\documentclass[prd,aps,superscriptaddress,nofootinbib,amsmath,amssymb,showpacs,9pt]{revtex4}
\usepackage{graphicx}
\usepackage{braket}
\usepackage{nicefrac}
\usepackage{float}%
\usepackage{amsmath}
\usepackage{hyperref}
\usepackage{multirow}
 \usepackage{bbold} 

\usepackage[usenames, dvipsnames]{color}

\def\slashchar#1{\setbox0=\hbox{$#1$}
   \dimen0=\wd0 \setbox1=\hbox{/} \dimen1=\wd1
   \ifdim\dimen0>\dimen1 \rlap{\hbox to \dimen0{\hfil/\hfil}} #1
   \else  \rlap{\hbox to \dimen1{\hfil$#1$\hfil}} / \fi}
\def\p{\slashchar{p}}
\def\q{\slashchar{q}}
\def\D{\slashchar{D}}

\graphicspath{{./plots/}}

\begin{document}
\title{Charged current (anti)neutrino induced eta production off the nucleon}

\author{A. \surname{Fatima}}
\affiliation{Department of Physics, Aligarh Muslim University, Aligarh-202 002, India}
\author{M. Sajjad \surname{Athar}}
\email{sajathar@gmail.com}
\affiliation{Department of Physics, Aligarh Muslim University, Aligarh-202 002, India}
\author{S. K. \surname{Singh}}
\affiliation{Department of Physics, Aligarh Muslim University, Aligarh-202 002, India}

\begin{abstract} 
The charged current (anti)neutrino induced eta production from the nucleons is studied in a model based on the effective Lagrangians to evaluate the contribution from the nonresonant and resonant diagrams. 
The contribution from the nonresonant background terms has been obtained 
using a microscopic model based on the SU(3) chiral Lagrangians. 
The contribution from the resonant diagrams due to the low lying $S_{11}(1535)$, $S_{11} (1650)$, and $P_{11} (1710)$ resonances has been evaluated using an effective phenomenological Lagrangian with its parameters determined from the experimental values of their branching ratios and decay widths to the $N\eta$ channel. 
The model is first used to reproduce satisfactorily the experimental data from the MAINZ and JLab on the total 
cross sections for the photo- and electro- production of $\eta$ mesons, which fixes the model parameters in the vector current interaction. 
The PCAC hypothesis and generalized Goldberger-Treiman relation are used to fix the parameters of the axial vector interaction.
The model is then applied to study the weak production of eta mesons induced by the neutrinos and antineutrinos, and predicts the numerical values for the $Q^2$-distribution, 
$\eta$-momentum distribution, and the total cross section for the reactions $\nu_{\mu} + n \longrightarrow \mu^{-} + p + \eta$ and $\bar{\nu}_{\mu} + p \longrightarrow \mu^+ + n + \eta$ in the energy region up to 2~GeV. 
It is found that the photo, electro, and  (anti)neutrino production of eta mesons is dominated by the contribution from the $S_{11}(1535)$ resonance.
The results discussed in this work are relevant for the present and future accelerator experiments like MicroBooNE, T2K, NOvA, MINERvA, T2-HyperK and DUNE as well as for the atmospheric neutrino experiments.  
\end{abstract}
\maketitle
\section{Introduction}
Many efforts are being made around the world to learn more about the neutrinos, with the top priorities focussed on obtaining the precise information about the mass
hierarchy of the neutrino mass states, and the CP violating phase delta~($\delta_{CP}$)~\cite{SajjadAthar:2021prg}. 
Various experiments using the accelerator and the atmospheric neutrinos are being carried out in the few GeV energy region for this purpose. All of these experiments make use of moderate to heavy nuclear targets such as carbon, oxygen, argon, iron, and lead, which 
necessitate an understanding of the nuclear medium effects that affects the (anti)neutrino-nucleon cross sections when the interaction takes place with the bound nucleons in these nuclei. 
The study of such nuclear medium effects makes use of a suitable model of the nucleus for describing the nuclear structure along with a good knowledge of the scattering amplitudes for the basic weak processes of (anti)neutrino scattering from the
free nucleons like the neutral current~(NC) elastic, charged current~(CC) quasielastic, inelastic, and deep inelastic scattering~(DIS). 
While the (anti)neutrino scattering processes of NC elastic, CC quasielastic and DIS from the free nucleons is well studied in the standard model, this is not the case with the inelastic scattering despite the enormous work done on the dominant process of the inelastic single pion production in the region of a few GeV. 
There are many other inelastic processes, in addition to the single pion production, like the production of kaons, hyperons with~(without) accompanying pions, multi pions, $\rho$, $\omega$ and $\eta$ mesons, which have not been studied with the same rigour as the single pion production. 
A study of these processes is important in the region of a few GeV energy where many higher resonances are excited which decay into these particles. 
Such studies contribute to a better understanding of the various electroweak processes of the excitation of higher resonances in the energy region of shallow inelastic scattering~(SIS) leading to the DIS region. 
The results in the SIS region and their comparison with the results in the DIS region provides an opportunity to understand the phenomenon of quark-hadron duality in electroweak interactions. 

In recent years, some work on the kaon and hyperon productions with and without accompanying pions have been done, but no work has been done on the $\eta$ production except a very early work by Dombey~\cite{Dombey:1968vh} and a recent work by Nakamura et al.~\cite{Nakamura:2015rta}. 
In view of the recent developments in the detector technology, it is possible to experimentally look for the weak production of 
$\eta$ mesons induced by neutrinos and antineutrinos in present experiments at MicroBooNE, T2K, NOvA, MINERvA, and future experiments at T2-HyperK and DUNE. For example, the MicroBooNE collaboration is currently performing the analysis to study the $\eta$ production cross section~\cite{talk_nuint}.
In addition to the additional weak processes of the $\eta$ production, which should be taken into account in modelling the neutrino cross sections being used in various neutrino event generators, the weak production of $\eta$ is an important source probe to search for the strangeness content of the nucleons~\cite{Dover:1990ic}. 
Moreover, the weak production of $\eta$ also plays an important role in the background studies in search of the proton decays through the $p \longrightarrow \eta e^+$ modes. In view of the topical importance of the weak production of $\eta$, we have studied in this work the CC production of $\eta$ from the free nucleons induced by the (anti)neutrinos through the reactions
\begin{eqnarray}
 \nu_{\mu} + n &\longrightarrow& \mu^{-} + p + \eta \\
 \bar{\nu}_{\mu} + p &\longrightarrow& \mu^+ + n + \eta
\end{eqnarray}
in a model using the effective Lagrangian approach. Separate effective Lagrangians are used to evaluate the contributions of the nonresonant Born terms and the resonant terms. 
An effective Lagrangian based on SU(3) chiral symmetry is used for the nonresonant 
Born terms. The basic parameters of the model are $f_m$, the meson decay constants, Cabibbo angle, 
the proton and neutron magnetic moments and the axial vector coupling constants 
for the baryon octet, $D$ and $F$,
 that are obtained from the analysis of the semileptonic decays of neutron and hyperons. 
 We consider low lying  $I= \frac12$ resonances such as $S_{11}$(1535), $S_{11}$(1650) and $P_{11}$(1710).
 The vector form factors of the $N-S_{11}$ and $N-P_{11}$ transitions have been obtained 
 from the helicity amplitudes extracted in the analysis of pion 
photo- and electro- production data using the MAID analysis in the unitary isobar model~\cite{Drechsel:2007if}.  
The present model has been earlier used to study the associated particle production 
using real photons off the proton target~\cite{Fatima:2020tyh}, weak interaction induced single pion production~\cite{RafiAlam:2015fcw}, kaon production~\cite{Alam:2013vwa, RafiAlam:2010kf, RafiAlam:2012bro}, two pion  production~\cite{Hernandez:2007ej}, etc. For a detailed description, please see Ref.~\cite{SajjadAthar:2022pjt}. 

Firstly, we have applied this model to obtain the results of the cross sections for 
 photo- and electro- induced eta production off the nucleon 
 target. 
 The production of $\eta$ particle in electromagnetic reactions induced by photons and electrons is well 
studied theoretically as well as experimentally~\cite{Benmerrouche:1991qx, Benmerrouche:1994uc, Tiator:1994et, Deutsch-Sauermann:1997bnp, Chiang:2001as, Feuster:1998cj, Davidson:1999in, Kirchbach:1996kw, Neumeier:2000fb, JeffersonLabE94014:1998czy, CLAS:2000mbw, Kamano:2019gtm, A2:2014pie, CrystalBallatMAMI:2010slt, Denizli:2007tq, Breitmoser:1996dy, Kuznetsov:2008hj, Crede:2009zzb, Ruic:2011wf, Fix:2003kq, Bartholomy:2007zz, Witthauer:2017wdb, Rosenthal:1991vh, GRAAL:2000qng, Price:1995vk, Krusche:1995nv} and the contribution of the vector contribution to these processes is fairly 
known.
 With the availability of high-duty cycle electron accelerators, such as the MIT-Bates,
ELSA at Bonn, MAMI at Mainz, NIKHEF at Watergraafsmeer, and CEBAF at JLab, efforts are being made both theoretically as well as 
experimentally to study the eta production induced by virtual as well as real photons from the nucleons and nuclei~\cite{Ruic:2011wf, Fix:2003kq, Bartholomy:2007zz, Crede:2009zzb}. 
Recently, results have been reported from MAMI-C~\cite{CrystalBallatMAMI:2010slt}
 using Crystal Ball and TAPS multi photon spectrometer in the energy range of 
707~MeV to 1.4~GeV for the differential as well as  the total scattering cross sections. 
We find that the theoretical results obtained in this explain well the photoproduction total cross section data from CLAS 2006 for $\gamma + p \rightarrow \Lambda + K$,
 and the photoproduction data from MAMI Crystal Ball~\cite{CrystalBallatMAMI:2010slt, A2:2014pie} for $\gamma + p \rightarrow p + \eta$ 
 and  $\gamma + n \rightarrow n + \eta$, and the electroproduction data
 from CLAS 2001 for 
 $e + p \rightarrow e + p + \eta$. These results have been used to fix the electromagnetic vector couplings and the $Q^2$ dependence of the vector $N-R$ transition form factors. 
 These informations are then used
 to obtain the isovector vector form factors for $N-R$ transitions.
 The axial vector coupling is fixed using the experimental
 $S_{11} \rightarrow N \pi$ and $P_{11} \rightarrow N \pi$ partial decay widths. 
 Assuming the pion-pole dominance 
together with the PCAC hypothesis, the pseudoscalar form factor is obtained in terms of the axial vector form factor. For the $Q^2$ dependence of the axial vector form factor, we have used a dipole form with the value of axial dipole mass taken to be  $M_{A}=1$~GeV. 

The plan of the paper is following. In sections-\ref{sec:eta:photo} and \ref{sec:eta:elctro}, we present the formalism for photon and electron induced eta production, respectively.
 In  section-\ref{sec:eta:weak}, the formalism for the charged current $\nu_l({\bar\nu}_l)$ induced eta production
has been presented. The results and discussions are presented in section-\ref{results} and section-\ref{conclude} concludes our findings.

\section{Formalism}
 The eta meson is an isoscalar pseudoscalar particle~($I=0,~J^P=0^-$), a member of the ground state SU(3) meson nonet with mass 547.86~MeV,
 a life time of about $5 \times 10^{-19}$sec and decays mainly to $3\pi~(\sim 56\%)$, $2\gamma~(\sim 39\%)$, and $\pi^+ \pi^- \gamma~(\sim 4.5\%)$ modes.
 The $\eta$ mesons are produced at $E_{\nu_l({\bar\nu}_l)} \ge 0.71~(0.88)$~GeV for $\nu_e~ ({\nu}_\mu)$ induced 
CC reactions. In the following sections, we have first applied our model to study the electromagnetic production of eta mesons induced by real and virtual photons, followed by the weak production of eta indued by (anti)neutrinos from the free nucleon target. 

\subsection{$\eta$ production induced by photons}\label{sec:eta:photo}

The differential cross section for the photoproduction of $\eta$ mesons off the free nucleon, {  i.e.},
\begin{equation}\label{eq:eta}
 \gamma(q) + N(p ) \longrightarrow N (p^{\prime}) + \eta(p_{\eta}),
\end{equation}
is written as
\begin{eqnarray}\label{eq:sigma_gen}
d\sigma &=& \frac{1}{4 (q\cdot p)} (2 \pi)^{4} \delta^{4}(q+p-p_{\eta}-p^{\prime}) \frac{d{\vec{p}_{\eta}}}{(2 \pi)^{3} (2 
E_{\eta})} \frac{d{\vec p\,}^{\prime}}{(2 \pi)^{3} (2 E^{\prime})} \overline{\sum_{r}} \sum | \mathcal M^{r} |^2,
\end{eqnarray}
where $N = p$ or $n$, the quantities in the parentheses of Eq.~(\ref{eq:eta}) represent the four momenta of the 
corresponding particles, $E_{\eta}$ and $E^{\prime}$, respectively, are the energies of the outgoing eta and nucleon. 
$ \overline{\sum} \sum | \mathcal M^{r} |^2$ is the square of the transition matrix element $\mathcal{M}^{r}$, for the photon 
polarization state $r$, averaged and summed over the initial and final spin states. 
\begin{figure} 
\begin{center}
\includegraphics[height=4.5cm,width=7cm]{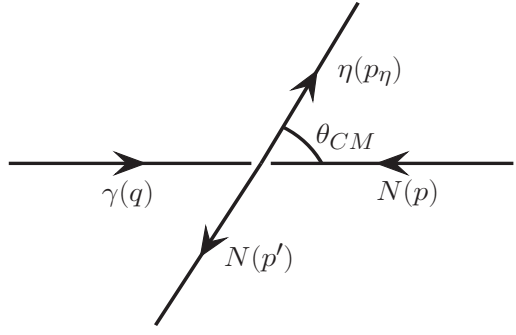}
\caption{Diagrammatic representation of the process $ \gamma (q) + N(p) \longrightarrow \eta(p_{\eta}) + N(p^{\prime})$ in the center of mass~(CM) frame. 
The quantities in the parentheses represent the four momenta of the corresponding particles. 
 $\theta^{CM}$ is the angle between photon and eta in the CM frame.  }\label{CM}
 \end{center}
 \end{figure}
 
Using Eq.~(\ref{eq:sigma_gen}), the differential cross section $\frac{d\sigma}{d\Omega}$ in the CM frame~(Fig.~\ref{CM}) is written as
\begin{equation}\label{dsig}
\left. \frac{d \sigma}{d \Omega}\right|_{CM} = \frac{1}{64 \pi^{2} s} \frac{|\vec{p}\;^{\prime}|}{|\vec{p}|} 
\overline{\sum_{r}} \sum_{spin} |\mathcal{M}^{r}|^2,
\end{equation}
where the CM energy $s$ is obtained as
\begin{equation}\label{s}
 s = W^2 = (q + p)^2 = M^{2} + 2M E_{\gamma} ,
\end{equation}
with $E_{\gamma}$ being the energy of the incoming photon in the laboratory frame. 

\begin{figure} 
\begin{center}
\includegraphics[height=8cm,width=0.95\textwidth]{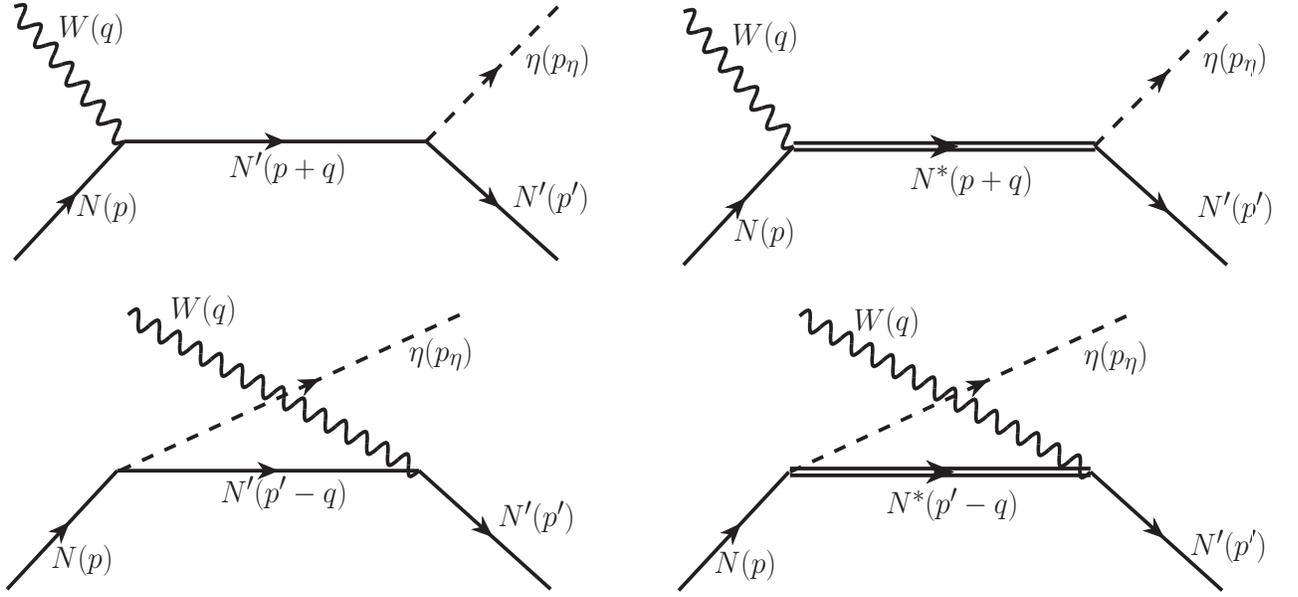}
\caption{Feynman diagrams corresponding to the nonresonant Born terms~(left panel) and resonance excitations~(right panel) for the process $ \gamma (q) + N(p) \longrightarrow \eta(p_{\eta}) + N(p^{\prime})$. Diagrams shown in the top panel are the nucleon pole diagrams, while the one shown in the bottom panel corresponds to the cross nucleon pole diagrams. In the case of electromagnetic interactions, $W=\gamma,\gamma^{\ast}$ and $N^{\prime} = N = p,n$, while in the case of CC induced weak interactions, $W=W^{\pm}$ and $N^{\prime}$ and $N$ corresponds to the different nucleons depending upon the charge conservation. 
The quantities in the parentheses represent the four momenta of the corresponding particles. }\label{Ch12_fg_eta:cc_weak_feynman}
 \end{center}
 \end{figure}

In the above expression, the transition matrix element $\mathcal{M}^{r}$ for the reaction~(\ref{eq:eta}) is written in terms of 
the real photon polarization vector $\epsilon_{\mu}^{r}$, {  i.e.}
\begin{equation}
\mathcal{M}^{r} = e \epsilon_{\mu}^{r} (q) \bra{N(p^{\prime}) \eta(p_{\eta})} {J}^{\mu} \ket{N(p)} ,
\end{equation}
where $e = \sqrt{4\pi \alpha}$ with $\alpha$ being the 
fine-structure constant, and $\epsilon_{\mu}^{r} (q)$ satisfy the condition
\begin{equation}\label{lep:photo}
 \sum_{r=\pm1} \epsilon^{*(r)}_{\mu}\epsilon^{(r)}_{\nu} \longrightarrow - g_{\mu \nu}. 
\end{equation}
The hadronic tensor 
${\cal J}^{\mu \nu}$ is defined in terms of the hadronic current $J^\mu$ as
\begin{equation}\label{had}
  {\cal J}^{\mu \nu} = \overline{\sum} \sum_{spins} {J^{\mu}}^{\dagger} J^{\nu} = \text{Tr} \left[(\p + M) \tilde 
  J^{\mu}({\p}^{\prime}+M^{\prime})J^\nu\right], \qquad \tilde J^\mu=\gamma_0(J^\mu)^{\dagger}\gamma_0,
\end{equation}
where $M$ and $M^{\prime}$ are the masses of the incoming and outgoing nucleons, respectively. The hadronic matrix element 
of the electromagnetic current $J^{\mu}$ receives contributions from the nonresonant background terms and the terms corresponding to 
the resonance excitations, shown in Fig.~\ref{Ch12_fg_eta:cc_weak_feynman} and discussed later in this section. 

Using Eqs.~(\ref{lep:photo}) and (\ref{had}), the transition matrix element squared is obtained as
\begin{equation}\label{mat}
\overline{\sum_{r}} \sum_{spin} |\mathcal{M}^{r}|^2 = -\frac{e^2}{4}g_{\mu \nu} {\cal J}^{\mu \nu},
\end{equation}
used for the cross sections defined in Eq.~(\ref{dsig}).
The hadronic currents for the various NR terms shown in Fig.~\ref{Ch12_fg_eta:cc_weak_feynman} are 
obtained using the nonlinear sigma model which in the following has been discussed very briefly.

\subsubsection{Nonlinear sigma model}\label{sec:sigma_model}
The lowest-order $SU(3)$ chiral Lagrangian in the nonlinear sigma model describing the pseudoscalar mesons in the presence of an external current, is given by~\cite{Scherer:2002tk, Scherer:2012xha}:
\begin{equation}\label{eq2:lagM}
{\cal L}_M=\frac{f_m}{4}\mbox{Tr}[D_\mu U (D^\mu U)^\dagger].
\end{equation}
 The covariant derivatives $D^{\mu} U$ and $D^{\mu} U^{\dagger}$ appearing 
in Eq.~(\ref{eq2:lagM}) are expressed in terms of the partial derivatives as
\begin{eqnarray}\label{eq2:coDer}
 D^\mu U \equiv \partial^\mu U - i r^\mu U + i U l^\mu, \qquad \qquad
  D^\mu U^\dagger \equiv \partial^\mu U^\dagger + i U^\dagger r^\mu - i l^\mu U^\dagger,
\end{eqnarray}
where $U$ is the $SU(3)$ unitary matrix given as
\begin{equation}
 U(x) = \exp\left(i\frac{\Phi(x)}{ f_m} \right), 
\end{equation}
where $f_m$ is the meson decay constant, $\Phi(x)$ corresponds to the $3 \times 3$ pseudoscalar meson matrix. $r_{\mu}$ and $l_{\mu}$, respectively, represent the right and 
left handed currents, defined in terms of the vector~($v_{\mu}$) and axial-vector~($a_{\mu}$) fields as
\begin{equation}\label{Ch11_NLSM_VA}
 l_{\mu} = \frac{1}{2}(v_{\mu} - a_{\mu}), \qquad \qquad r_{\mu} = \frac{1}{2}(v_{\mu} + a_{\mu}).
\end{equation}
The vector and axial-vector fields are different for the interaction of the different gauge boson fields with the meson fields. 

In the case of electromagnetic gauge fields, the left and right handed currents are identical and are expressed as
\begin{equation}\label{Ch11_lr_EM}
 l_{\mu} = r_{\mu} = -e \hat{Q} A_{\mu} ,
\end{equation}
where $e$ is the strength of the electromagnetic interaction, $A_{\mu}$ represents the photon field and $\hat{Q}=\begin{pmatrix}
                                                                                      2/3 & 0 & 0 \\
                                                                                      0 & -1/3 & 0 \\
                                                                                      0 & 0 & -1/3
                                                                                      \end{pmatrix}$ represents the charge 
                                                                                      of the
                                                                                      $u,~ 
d$, and $s$ quarks. In the case of weak CC induced processes, the left and right handed currents are expressed as
\begin{equation}\label{Ch11_lr_CC}
 l_{\mu} = - \frac{g}{2} (W_{\mu}^{+} T_{+} + W_{\mu}^{-}T_{-}), \qquad \qquad r_{\mu} = 0,
\end{equation}
where $g = \frac{e}{\sin \theta_{W}}$, $\theta_{W}$ is the Weinberg angle, $W_{\mu}^{\pm}$ represents the W-boson field 
and $T_{\pm}$ is defined as
\begin{equation}
T_{+} = \begin{pmatrix}
        0 & V_{ud} & V_{us} \\
        0 & 0 & 0 \\
        0 & 0 & 0
        \end{pmatrix}, \qquad \text{and} \qquad T_{-} = \begin{pmatrix}
                                                        0 & 0 & 0 \\
                                                        V_{ud} & 0 & 0 \\
                                                        V_{us} & 0 & 0
        \end{pmatrix},
\end{equation}
with $V_{ud} = \cos \theta_{C}$ and $V_{us} = \sin \theta_{C}$ being the elements of the Cabibbo-Kobayashi-Maskawa matrix 
and $\theta_{C}$ being the Cabibbo angle.

The lowest-order chiral Lagrangian for the baryon octet in the presence of an external current, may be written in terms of 
the $SU(3)$ matrix of the baryons $B$ as~\cite{Scherer:2002tk, Scherer:2012xha},
\begin{equation}\label{eq2:lagB}
{\cal L}_{MB}=\mbox{Tr}\left[\bar{B}\left(i\D
-M\right)B\right]
-\frac{D}{2}\mbox{Tr}\left(\bar{B}\gamma^\mu\gamma_5\{u_\mu,B\}\right)
-\frac{F}{2}\mbox{Tr}\left(\bar{B}\gamma^\mu\gamma_5[u_\mu,B]\right),
\end{equation}
where $M$ denotes the mass of the baryon octet, $D=0.804$ and $F=0.463$ are the symmetric and antisymmetric axial-vector 
coupling constants for the baryon octet,  and the Lorentz vector $ u^\mu$ is given by~\cite{Scherer:2012xha}:
\begin{equation}\label{eq2:vielbein}
u^\mu = i \left[ u^\dagger ( \partial^\mu - i r^\mu) u - u ( \partial^\mu - i l^\mu) u^\dagger \right].
\end{equation} 
In the case of meson-baryon interactions, the unitary matrix for the pseudoscalar field is expressed as 
\begin{equation}
 u = \sqrt U \equiv \exp \left( i \frac{\Phi(x)}{ 2 f_m } \right),
\end{equation}
and the covariant derivative $D_{\mu}$ on the baryon fields $B$ is given by
\begin{equation}\label{dmuB}
D_\mu B=\partial_\mu B +[\Gamma_\mu,B], \qquad \text{with} \qquad \Gamma^\mu=\frac{1}{2}\left[u^\dagger(\partial^\mu-
ir^\mu)u
+u(\partial^\mu-il^\mu)u^\dagger\right],
\end{equation}
which is known as the chiral connection.

\subsubsection{Currents for the nonresonant Born terms and resonance excitations}
Expanding the Lagrangians given in Eqs.~(\ref{eq2:lagM}) and (\ref{eq2:lagB}) for the lowest lying baryons and mesons, one obtains the Lagrangians for any desired vertex involving the interactions of mesons and baryons among themselves or with the external ﬁelds. Using these Lagrangians, 
the expressions of the hadronic currents for $s$-, 
$u$- channels of the $\eta$ photoproduction processes, corresponding to the Feynman diagrams shown in  Fig.~\ref{Ch12_fg_eta:cc_weak_feynman}~(left panel), are obtained as~\cite{SajjadAthar:2022pjt}:
\begin{eqnarray}\label{j:s}
J^\mu \arrowvert_{sN} &=&- A_{s}~F_{s}(s) \bar u(p^\prime) \p_{\eta} \gamma_5 \frac{\p + \q + M}
  {s -M^2} \left(\gamma^\mu e_{N} +i \frac{\kappa_{N}}{2 M} \sigma^{\mu \nu} q_\nu \right) u(p), \\
  \label{j:ulam}
J^\mu \arrowvert_{uN} &=&- A_{u} ~F_{u} (u) \bar u(p^\prime) \left(\gamma^\mu e_{N} + i \frac{\kappa_{N}}{2 M} \sigma^{\mu 
\nu} q_\nu \right) \frac{ \p^{\prime} - \q + M}{u - M^2} \p_\eta \gamma_5 u(p),
\end{eqnarray}
where $N$ stands for a proton or a neutron in the initial and final states, $s$ is defined in Eq.~(\ref{s}) and $u = 
(p^{\prime} - q)^{2}$,
$A_{i}$'s; $i=s,u$ are the coupling strengths of $s$, and $u$ channels, respectively, and are obtained as~\cite{SajjadAthar:2022pjt}
\begin{eqnarray}\label{eq:coupling}
 A_{s} = A_{u} &=& \left(\frac{D - 3F}{2 \sqrt{3} f_{\eta}}\right) .
 \end{eqnarray}
$D$ and $F$ are the axial-vector couplings of the baryon octet and $f_{\eta}=105$~MeV~\cite{Faessler:2008ix} is the $\eta$ decay constant. 
The value of $\kappa$ for proton, and neutron are $\kappa_{p} = 1.7928$, and $\kappa_{n} = -1.913$ in units of $\mu_{N}$~\cite{ParticleDataGroup:2020ssz}. 

In order to take into account the hadronic structure of the nucleons, the form factors $F_{s} (s)$, and $F_{u} (u)$, are 
introduced at the strong vertex. Various parameterizations of these form factors are available in the 
literature~\cite{Skoupil:2016ast}. We use the most general form of the hadronic form factor which is 
taken to be of the dipole form~\cite{Fatima:2020tyh}:
\begin{equation}\label{FF_Born}
F_{x} (x) = \frac{\Lambda_{B}^{4}}{\Lambda_{B}^{4} + (x - M_{x}^{2})^{2}}, \qquad \qquad \quad x=s,u
\end{equation}
where $\Lambda_{B}$ is the cut-off parameter taken to be the same for the s- and u-channel nonresonant Born terms.  $x$ represents the Mandelstam variables $s,~u$, and $M_{x} = M$ 
corresponds to the mass of the exchanged nucleons in the $s$ and $u$ channels. The value of $\Lambda_{B}$ 
is fitted to the experimental data for both the proton and neutron targets simultaneously and the best fitted value is 
$\Lambda_{B}=0.78$~GeV for $s$- and $u$-channel diagrams.
One of the most important property of the electromagnetic current is the gauge invariance which ensures the current 
conservation and is implemented in the case of $\eta$ production.

In the present work, we have taken into account only the low lying resonances, which have mass $M_{R}<1.8$~GeV and have a significant branching ratio to the $N\eta$ mode reported in PDG~\cite{ParticleDataGroup:2020ssz}. Specifically, we have considered three spin $\frac{1}{2}$ resonances {  viz.} $S_{11} 
(1535)$, $S_{11} (1650)$, and $P_{11} (1710)$. The general properties of these resonances like mass, decay width, spin, etc. are given in Table~\ref{tab:param-p2}, where we see that $S_{11}(1535)$ resonance dominates the coupling to the $N\eta$ channel.

The hadronic vector transition current for the spin $\frac12$ resonance state is given by~\cite{Athar:2020kqn}  
\begin{eqnarray}\label{had_curr_1/2}
j^{\mu}_{\frac{1}{2}}=\bar{u}(p') \Gamma^\mu_{\frac12} u(p), 
\end{eqnarray}
where $u(p)$ and $\bar u(p^\prime)$ are, respectively, the Dirac spinor and the adjoint Dirac spinor for spin $\frac{1}{2}$ 
particles and $\Gamma^\mu_\frac12$ is the vertex function. For a positive parity state, $\Gamma^{\mu}_{\frac{1}{2}^+}$ is given 
by 
\begin{align}\label{eq:vec_half_pos}
  \Gamma^{\mu}_{\frac{1}{2}^+} &= {V}^{\mu}_\frac{1}{2},
  \end{align}
and for a negative parity resonance, $\Gamma^{\mu}_{\frac{1}{2}^-}$ is given by  
\begin{align}\label{eq:vec_half_neg}
  \Gamma^{\mu}_{\frac{1}{2}^-} &= {V}^{\mu}_\frac{1}{2} \gamma_5 ,
  \end{align}
where $V^{\mu}_{\frac{1}{2}}$ represents the vector current parameterized in terms of $F_{2}^{R^{+},R^{0}} $, as
 \begin{align}\label{eq:vectorspinhalf1}
  V^{\mu}_{\frac{1}{2}} & =\left[\frac{F_2^{R^{+},R^{0}}}{2 M} 
  i \sigma^{\mu\alpha} q_\alpha \right].
\end{align}
The coupling $F^{R^{+},R^0}_{2}$ is derived from the helicity amplitudes extracted from the real photon scattering 
experiments. The explicit relation between the coupling $F_2^{R^{+},R^0}$ and the helicity amplitude $A_{\frac{1}{2}}^{p}$ is given 
by~\cite{Fatima:2020tyh}:
\begin{eqnarray}\label{eq:hel_spin_12}
A_\frac{1}{2}^{p,n}&=& \sqrt{\frac{2 \pi \alpha}{M} \frac{(M_R \mp M)^2}{M_R^2 - M^2}} \left[ \frac{M_R \pm M}{2 M} F_2^{R^{+},R^0} 
\right] ,
\end{eqnarray}
where the upper~(lower) sign stands for the positive~(negative) parity resonance. $R^{+}$ and $R^{0}$ correspond, respectively, to the charged and neutral states of the isospin $\frac{1}{2}$ resonances. $M_R$ is the mass of corresponding resonance. 
The value of the helicity amplitude $A_{\frac{1}{2}}^{p,n}$ for the $S_{11}(1650)$ resonance is taken from MAID~\cite{Tiator:2011pw}, 
while for the other spin $\frac{1}{2}$ nucleon resonances, these values are taken from PDG~\cite{ParticleDataGroup:2020ssz} and are quoted in 
Table~\ref{tab:resonance}.

\begin{table}
\centering
\begin{tabular*}{140mm}{@{\extracolsep{\fill}}ccccccc}\hline \hline
&\multicolumn{2}{c}{Resonance $\rightarrow$} & \multirow{2}{*}{$S_{11}(1535)$} & \multirow{2}{*}{$S_{11}(1650)$} & \multirow{2}{*}{$P_{11} (1710)$} & \\
&\multicolumn{2}{c}{Parameters $\downarrow$} &&& &\\ \hline
&\multicolumn{2}{c}{$M_{R}$ (GeV)} & $1.510 \pm 0.01$ & $1.655 \pm 0.015$ & $1.700 \pm 0.02$ &\\ \hline
&\multicolumn{2}{c}{$\Gamma_{R}$ (GeV)} & $0.130 \pm 0.02$ & $0.135 \pm 0.035$ & $0.120 \pm 0.04$ &\\ \hline
&\multicolumn{2}{c}{$I(J^P)$} &$\frac{1}{2}(\frac{1}{2}^{-})$& $\frac{1}{2}(\frac{1}{2}^{-})$ & $\frac{1}{2}(\frac{1}{2}^{+})$ &\\ \hline 
&\multirow{4}{*}{Branching ratio (in \%)} & $N\pi$ & $32-52$~(43)& $50-70$~(60) & $5-20$~(16)&\\ 
&& $N\eta$ & $30-55$~(40) & $15-35$~(25) & $10-50$~(20) &\\ 
& &$K\Lambda$ &$-$& $5-15$~(10) & $5-25$~(15) &\\ 
&& $N\pi\pi$ &$4-11$~(17)& $20-58$~(5)& $14-48$~(49)&\\ \hline
&\multicolumn{2}{c}{$|g_{RN\pi}|$} & 0.1019 & 0.0915 & 0.0418 &\\ \hline
&\multicolumn{2}{c}{$|g_{RN\eta}|$} & 0.3696 & 0.1481 & 0.1567 &\\ \hline \hline
\end{tabular*}
\caption{Properties of the spin $\frac{1}{2}$ resonances available in the PDG~\cite{ParticleDataGroup:2020ssz}, with Breit-Wigner mass $M_{R}$, the total decay width $\Gamma_{R}$,
isospin $I$, spin $J$, parity $P$, the branching ratio full range available from PDG~(used in the present calculations) into different meson-baryon like $N\pi$, $N\eta$, $K\Lambda$, and $N\pi\pi$, and the strong coupling constant $g_{RN\pi}$ and $g_{RN\eta}$.}
\label{tab:param-p2}
\end{table}

%

The most general form of the hadronic currents for the $s-$ and $u-$ channel processes where a resonance state $R_{\frac12}$ 
is produced and decays to a $\eta$ and a nucleon in the final state, are written as
\begin{eqnarray}
j^\mu\big|_{s}&=&  \frac{g_{RN\eta}}{f_{\eta}} \bar u({p}\,') 
 \p_{\eta} \Gamma_{s} \left( \frac{\p+\q+M_{R}}{s-M_{R}^2+ iM_{R} \Gamma_{R}}\right) 
 \Gamma^\mu_{\frac12 
 \pm} u({p}\,), \nonumber\\
 \label{eq:res1/2_had_current}
 j^\mu\big|_{u}&=&  \frac{g_{RN\eta}}{f_{\eta}} \bar u({p}\,') 
 \Gamma^\mu_{\frac12 \pm}\left(\frac{\p^{\prime}-\q+M_{R}}{u-M_{R}^2+ iM_{R} \Gamma_{R}}\right) 
 \p_{\eta} \Gamma_{s}  u({p}\,),
\end{eqnarray}
where $\Gamma_{R}$ is the decay width of the resonance, $\Gamma_{s} = 1(\gamma_{5})$ stands for the positive~(negative) 
parity resonances. $\Gamma_{\frac{1}{2}^{+}}$ and $\Gamma_{\frac{1}{2}^{-}}$ are, respectively, the vertex functions for the 
positive and negative parity resonances, defined in Eqs.~(\ref{eq:vec_half_pos}) and (\ref{eq:vec_half_neg}), respectively. 
$g_{RN\eta}$ is the coupling strength for the process $ R \to N\eta$, given in Table~\ref{tab:param-p2}. 

We determine the $RN\eta$ coupling using the value of branching ratio and decay width of these 
resonances from PDG~\cite{ParticleDataGroup:2020ssz} and use the expression for the decay rate which is obtained by writing the most general form of 
$RN\eta$ Lagrangian~\cite{Leitner:2008ue}:
\begin{align}\label{eq:spin12_lag}
 \mathcal{L}_{R N\eta} &= \frac{g_{ R N\eta} }{f_{\eta}}\bar{\Psi}_{R} \; 
 \Gamma^{\mu}_{s} \;
  \partial_\mu \eta^i \tau_i \,\Psi ,
\end{align}
where $g_{RN\eta}$ is the $RN\eta$ coupling strength. $\Psi$ is the nucleon field and ${\Psi}_{R}$ 
is the resonance field. $\eta^i$ is the eta field and $\tau$ is the isospin factor for the isospin $\frac12$ 
states. The interaction vertex $\Gamma^{\mu}_{s} = \gamma^\mu \gamma^5$~($\gamma^\mu$) stands for positive~(negative) parity 
resonance states. 

Using the above Lagrangian, one obtains the expression for the decay width in the resonance rest frame as~\cite{SajjadAthar:2022pjt}:
\begin{align}\label{eq:12_width}
 \Gamma_{R \rightarrow N\eta} &= \frac{\mathcal{C}}{4\pi} \left(\frac{g_{RN\eta }}{f_{\eta}}
 \right)^2 \left(M_R \pm M\right)^2 \frac{E_{N} \mp M}{M_R} |\vec{p}^{\,\mathrm{cm}}_{\eta}|,
\end{align}
where the upper~(lower) sign represents the positive~(negative) parity resonance. The parameter $\mathcal{C}$ depends upon the 
charged state of $R$, $N\eta$ and is obtained from the isospin analysis and found out to be $1$. $|\vec p^{\,cm}_{\eta}|$ is the 
outgoing eta momentum measured from the resonance rest frame and is given by, 
\begin{equation}\label{eq:pi_mom}
|\vec{p}^{\,\mathrm{cm}}_{\eta}| = \frac{\sqrt{(W^2-M_{\eta}^2-M^2)^2 - 4 M_{\eta}^2 M^2}}{2 M_R}  
\end{equation}
and $E_N$, the outgoing nucleon energy is
\begin{equation}\label{eq:elam}
  E_N=\frac{W^2+M^2-M_{\eta}^2}{2 M_R},
\end{equation}
where $W$ is the total center of mass energy carried by the resonance.

 \begin{figure}  
\begin{center}
\includegraphics[width=0.85\textwidth,height=8cm]{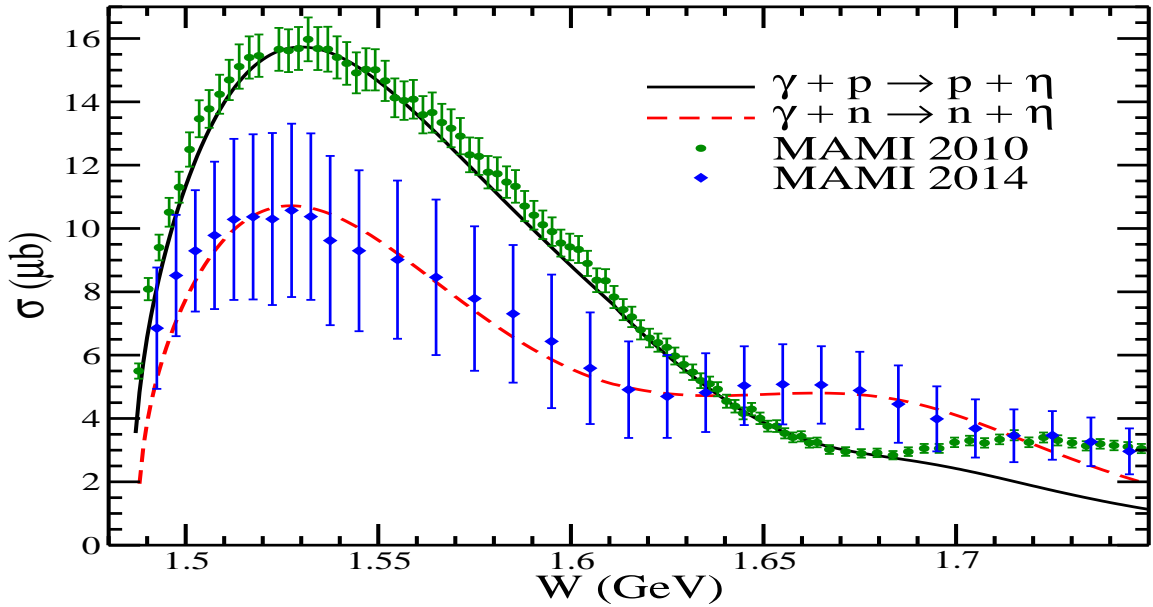}
\caption{Total cross section $\sigma$ vs. $W$ for $\gamma  p \longrightarrow \eta p$~(solid line) and $\gamma  n \longrightarrow \eta n$~(dashed line) processes. The experimental points for the proton target~(solid circle) are 
obtained from MAMI crystal ball collaboration~\cite{CrystalBallatMAMI:2010slt}, and for the neutron target~(solid diamond) we have used the quasifree neutron data from 
MAMI A2 collaboration~\cite{A2:2014pie}. }
\label{fg_eta:photo_xsec_mami}
\end{center}
\end{figure}
In analogy with the NR terms,  we have considered the following form factors at the strong 
vertex, in order to take into account the hadronic structure:
\begin{equation}
F^{*}_{x} (x) = \frac{\Lambda_{R}^{4}}{\Lambda_{R}^{4} + (x - M_{x}^{2})^{2}},
\end{equation}
where $\Lambda_{R}$ is the cut-off parameter whose value is fitted to the experimental data, $x$ represents the Mandelstam 
variables $s,~u$, and $M_{x} = M_{R}$ corresponding to the mass of the nucleon resonances exchanged in the $s,$ and $u$ 
channels. In general, $\Lambda_{R}$ would be different from $\Lambda_{B}$, however, in the case of $\eta$ production by 
photons, it happens that the same value of $\Lambda_{R}$ as that of $\Lambda_{B}$ i.e. $\Lambda_{R} = \Lambda_{B} =
0.78$~GeV gives the best fit to the experimental data. The same values of $\Lambda_{R}$ and $\Lambda_{B}$ help us to minimize the number of free parameters used to fit 
the experimental data.

In Fig.~\ref{fg_eta:photo_xsec_mami}, we have presented the results for the total scattering cross section $\sigma$ as a 
function of $W$ for $\gamma + p \longrightarrow p + \eta$ and $\gamma + n \longrightarrow n + \eta$ 
processes in the region of $W$ from $\eta$ production threshold to $K\Lambda$ production threshold. We have compared our 
theoretical results with the experimental data obtained by McNicoll et al.~\cite{CrystalBallatMAMI:2010slt}  for the MAMI crystal ball
collaboration on the proton target and the quasifree neutron data from Werthmuller et 
al.~\cite{A2:2014pie} for the MAMI A2 collaboration. It may be observed from the figure that in the case of $\eta$ production from the proton and neutron 
targets, our results, with a very few free parameters, are in a very good agreement with the available experimental data. 

\begin{figure}  
\begin{center}
\includegraphics[width=0.85\textwidth,height=6.5cm]{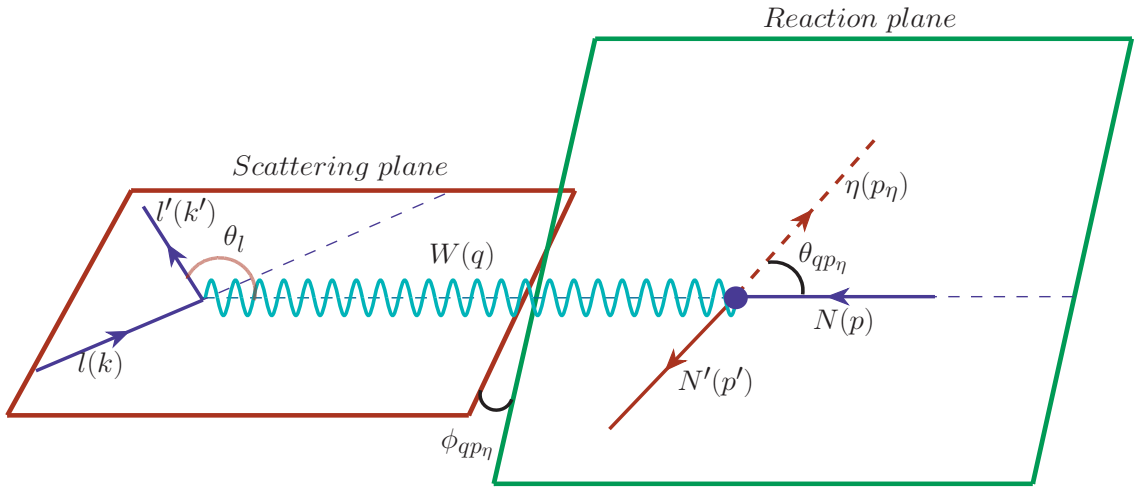}
\caption{Electron~(electromagnetic)/(anti)neutrino~(weak) scattering and reaction planes, depicting the hadronic plane in CM frame and scattering plane in the laboratory
frame. The kinematical variables used in the calculation of the scattering cross sections are defined in the figure.}
\label{fg_3D}
\end{center}
\end{figure}
\subsection{Electroproduction of eta meson}\label{sec:eta:elctro}
The electron induced $\eta$ production off the nucleon target is given by the reaction
\begin{equation} \label{eq:elecprod}
  e^- (k) + N(p) \longrightarrow e^-(k^\prime)  + N(p^{\prime}) + \eta (p_{\eta}) \,,
\end{equation}
where the four-momentum for each particle is indicated in the parentheses.
The four-momentum of the virtual photon exchanged in electroproduction is
given by $q = k - k^\prime$. 

The
differential cross section for the electroproduction process can be written
as
\begin{eqnarray}\label{eq:inelastic:reaction}
d\sigma &=& \frac{1}{4 ME_e(2\pi)^{5}} \frac{d{\vec k}^{\prime}}{ (2 E_{l})} 
\frac{d{\vec p\,}^{\prime}}{(2 E_{N})} \frac{d{\vec p}_{\eta}}{ (2 E_\eta)}
 \delta^{4}(k+p-k^{\prime}-p^{\prime}-p_{\eta})\overline{\sum}\sum | \mathcal M |^2,\;\;\;\;\;
\end{eqnarray}
where $ k( k^\prime) $ is the four momentum of the incoming~(outgoing) electron with energy $E_e( E_l)$; $p$ is 
the four momentum of the incoming nucleon which is at rest, $E_N$ and $p^\prime$ are respectively the energy and four 
momentum of the outgoing nucleon, and the four momentum of $\eta$ is $p_\eta$ with energy $ E_\eta$, and $M$ is the nucleon mass. The 
different kinematical variables used in the numerical calculations of the scattering cross section are depicted in 
Fig.~\ref{fg_3D}, where the scattering plane is in the laboratory frame while the reaction plane is in the center 
of mass~(CM) frame. $\overline{\sum}\sum | \mathcal M |^2  $ is the square of the transition amplitude averaged~(summed) over the 
spins of the initial~(final) states  and the transition matrix element is written in terms of the leptonic and the hadronic 
currents as 
\begin{equation}
\label{eq:Gg}
 \mathcal M = \frac{e^2}{{q^2}}\, {l_\mu} j^{\mu},
\end{equation}
where $l_{\mu}$ and $j^{\mu}$, respectively, are the leptonic and hadronic currents. The leptonic current is given as
\begin{equation}\label{eq:lepton_current}
 l_{\mu} = \bar{u} (k^{\prime}) \gamma_{\mu} u(k),
\end{equation}
and $j^{\mu}$ is the sum of the hadronic currents corresponding to the Born terms and resonance excitations, which will be discussed later in Section~\ref{current_em}.

Integrating over the three momentum of the outgoing nucleon, the expression for the differential scattering cross section 
given in Eq.~(\ref{eq:inelastic:reaction}) in the hadronic CM frame becomes
\begin{equation}\label{dsigma:pion}
\frac{d^5 \sigma}{dE_{l} ~d \Omega_{l} d\Omega_{qp_\eta}} = \frac{1}{32(2\pi)^{5}} \frac{|\vec{k}^{\prime}| |\vec{p}_{\eta}|}{E_{e}M W} \overline{\sum}\sum | \mathcal M |^2.
\end{equation}
The five-fold differential cross section for the electroproduction can also be
expressed as~\cite{Donnachie:1978fm, Amaldi:1979vh, Drechsel:1994zx}:
\begin{equation} \label{eq:5fdxs}
 \frac{d\sigma}{d\Omega_l\, dE_l\, d\Omega_{qp_\eta}}
 = \Gamma\, \frac{d\sigma_\text{v}}{d\Omega_{qp_{\eta}}} \,,
\end{equation}
with the flux of the virtual photon field given by
\begin{equation}
 \Gamma = \frac{\alpha}{2 \pi^2}\, \frac{E_l}{E_e}\,
          \frac{K}{Q^2}\, \frac{1}{1-\varepsilon} \,.
\end{equation}
In the above equation, $K =(W^2 - M^2) / 2 M$
denotes the ``photon equivalent energy'', the laboratory energy necessary
for a real photon to excite a hadronic system with CM energy $W$. The transverse polarization parameter of the virtual photon
\begin{equation}
 \varepsilon = \left(1 + 2 \frac{|\vec{q}|^2}{\,Q^2}\tan^2\frac{\theta_l}{2}
               \right)^{-1} \,,
\end{equation}
where $Q^2 = - q^2 = -(k-k^\prime)^2$.

It is
useful to express the angular distribution of the eta mesons in the CM
frame of the final hadronic states, particularly for the use of multipole
decompositions. Therefore, the virtual photon cross section $d\sigma_\text{v} / d\Omega$ should be evaluated in the CM frame, while the five-fold
differential cross section in Eq.~(\ref{eq:5fdxs}) is interpreted with
the flux factor in the lab frame.
 By choosing the energies of the
initial and final electrons and the scattering angle $\theta_l$ (see
Fig.~\ref{fg_3D}), we can fix the momentum transfer $Q^2$ and the
polarization parameter $\varepsilon$ of the virtual photon.

%

\subsubsection{Currents for the nonresonant Born terms and resonance excitations}\label{current_em}
Currents corresponding to the nucleon Born terms for the electroproduction of eta mesons, depicted in Fig.~\ref{Ch12_fg_eta:cc_weak_feynman}, are obtained using the nonlinear sigma model discussed in Section~\ref{sec:sigma_model} and are  written as: 
\begin{eqnarray}\label{Eq_eta:amp_photo}
J_{N(s)}^\mu &=&  
\frac{D-3F}{2\sqrt3 f_\eta} \bar u_N (p^\prime) \slashchar{p_\eta} \gamma^5  
\frac{\slashchar{p}+\slashchar{q}+M}{(p+q)^2-M^2} 
{\cal O}^\mu_N u_N (p) \nonumber \\ 
J_{N(u)}^\mu &=&   \frac{D-3F}{2\sqrt3 f_\eta} 
\bar u_N (p^\prime) {\cal O}^\mu_N
  \frac{\slashchar{p}-\slashchar{p}_{\eta}+M}{(p - p_\eta)^2-M^2} 
\slashchar{p}_{\eta} \gamma^5 u_N (p),
\end{eqnarray}
where the 
$\gamma N N$ vertex operator ${\cal O}^\mu_N$ is expressed in terms of the $Q^2$ dependent nucleon form factors as,
\begin{eqnarray}\label{eq:gNN_vertex}
{\cal O}^\mu_N &=& F_1^{N}(Q^2)\gamma^\mu + F_2^{N}(Q^2) i \sigma^{\mu\nu} 
\frac{q_\nu}{2M} .
\end{eqnarray}
The Dirac and Pauli form factors of the nucleon viz. $F_{1}^{p,n}(Q^2)$ and $F_{2}^{p,n} (Q^2)$, respectively, may be expressed in terms of the 
Sach's electric~($G_E^{p,n}(Q^2)$) and magnetic~($G_M^{p,n}(Q^2)$) form factors of the nucleons   as,
\begin{eqnarray}\label{Eq_eta:f1pn_f2pn}
F_1^{p,n}(Q^2)&=&\left(1+\frac{Q^2}{4M^2}\right)^{-1}~\left[G_E^{p,n}(Q^2)+\frac
{Q^2}{4M^2}~G_M^{p,n}(Q^2)\right] \nonumber \\
F_2^{p,n}(Q^2)&=&\left(1+\frac{Q^2}{4M^2}\right)^{-1}~\left[G_M^{p,n}(Q^2)-G_E^{
p,n}(Q^2)\right]. 
\end{eqnarray}
There are various parameterization for 
$G_{E,M}^{p,n}(Q^2)$ those are available in the literature. For the present work we have taken 
 the parameterization of these form factors from Bradford et al.~\cite{Bradford:2006yz} also known as BBBA05 parameterization.

Now we discuss the hadronic current corresponding to the resonance excitations and their subsequent decay to $N\eta$ channel.  The general expression of the hadronic current for the resonance excitation in the s- and u- channels, corresponding to the Feynman diagrams shown in Fig.~\ref{Ch12_fg_eta:cc_weak_feynman}~(right panel), are written, as,
\begin{eqnarray}\label{jhad:eta_electro}
j^\mu\big|_{s}&=&  \frac{g_{RN\eta}}{f_{\eta}} \bar u({p}\,') 
 \p_{\eta} \Gamma_{s} \left( \frac{\p+\q+M_{R}}{s-M_{R}^2+ iM_{R} \Gamma_{R}}\right) 
 \Gamma^\mu_{\frac12 
 \pm} u({p}\,), \nonumber\\
 j^\mu\big|_{u}&=&  \frac{g_{RN\eta}}{f_{\eta}} \bar u({p}\,') 
 \Gamma^\mu_{\frac12 \pm}\left(\frac{\p^{\prime}-\q+M_{R}}{u-M_{R}^2+ iM_{R} \Gamma_{R}}\right) 
 \p_{\eta} \Gamma_{s}  u({p}\,),
\end{eqnarray}
where $g_{RN\eta}$ is the strong coupling constant, which we have fixed using the $\eta$ production data. The vertex function $\Gamma^\mu_{\frac12 \pm}$ for the positive and negative parity resonances is given in Eqs.~(\ref{eq:vec_half_pos}) and (\ref{eq:vec_half_neg}), respectively, where the vector current $V_{\frac{1}{2}}^{\mu}$ in the case of electroproduction processes is expressed in terms of the $Q^2$ dependent form factors $F^{R^+,R^0}_{1,2} (Q^2)$ as
\begin{eqnarray}\label{eq_eta:nstar_em_vertex}
V_{\frac{1}{2}}^{\mu} &=& \frac{F_1^{R}(Q^2)}{(2 M)^2}(\slashchar{q} q^\mu+Q^2\gamma^\mu) 
+ \frac{F_2^{R}(Q^2)}{2 M} i \sigma^{\mu\nu} q_\nu , \qquad R=R^+,R^0  .
\end{eqnarray} 
The electromagnetic transition form factors  for the charged~($F_{1,2}^{R^+}(Q^2)$)  and neutral~($F_{1,2}^{R^0} (Q^2)$) states are then  related to the helicity amplitudes given by the following relations~\cite{SajjadAthar:2022pjt}; 
\begin{eqnarray}\label{eq2}
 A_{\frac{1}{2}}&=&\sqrt{\frac{2\pi\alpha}{K_R}} 
 \Bra{R,J_Z=\frac{1}{2}}\epsilon_\mu^{+} J_i^\mu \Ket{N,J_Z=\frac{-1}{2}}\zeta \nonumber \\ 
 S_{\frac{1}{2}}&=&-\sqrt{\frac{2\pi\alpha}{K_R}}\frac{|\vec q|}{\sqrt{Q^2}} 
\Bra{R,J_Z=\frac{1}{2}}\epsilon_\mu^{0} J_i^\mu \Ket{N,J_Z=\frac{-1}{2}}\zeta
\end{eqnarray}
where in the resonance rest frame, 
\begin{eqnarray}\label{eq1}
K_R&=&\frac{M_R^{2}-M^2}{2M_R}, \quad \qquad |\vec q |^2=\frac{(M_R^{2}-M^{2}-Q^{2})^2}{4M_R^{2}}+Q^2,\nonumber \\
\epsilon^\mu_{\pm}&=&\mp\frac{1}{\sqrt{2}}(0,1,\pm i,0),\qquad \qquad
\epsilon^\mu_{0}=\frac{1}{\sqrt{Q^{2}}}(|\vec q|,1,0 ,q^0).
\end{eqnarray}
The parameter $ \zeta $ is model dependent which is related to the sign of $R \rightarrow N \pi$, 
and for the present calculation is taken as $\zeta =1$. 

\begin{table}
\centering
\begin{tabular*}{160mm}{@{\extracolsep{\fill}}ccc c c c c  c c c}\hline\hline
&Resonance & Helicity amplitude & \multicolumn{3}{c}{ Proton target } & \multicolumn{3}{c}{ Neutron target } &\\ \hline
&&& ${\cal A}_{\alpha} (0)$ & $a_1$ & $b_{1}$ & ${\cal A}_{\alpha} (0)$ & $a_1$ & $b_{1}$ &\\ \hline
&\multirow{2}{*}{$S_{11}(1535)$} & $A_{\frac{1}{2}}$ & 95.0 & 0.5 & 0.51 & $-78.0$ & 1.75 & 1.75& \\ 
&&$ S_{\frac{1}{2}}$ & $-2.0$ & 23.9 & 0.81 & $32.5$ & 0.4 & 1.0& \\ \hline
&\multirow{2}{*}{$S_{11}(1650)$} & $A_{\frac{1}{2}}$ & 33.3 & 1.45 & 0.62 & $26.0$ & 0.1 & 2.5& \\ 
&&$ S_{\frac{1}{2}}$ & $-3.5$ & 2.88 & 0.76 & $3.8$ & 0.4 & 0.71& \\ \hline
&\multirow{2}{*}{$P_{11}(1710)$} & $A_{\frac{1}{2}}$ & 50.0 & 1.4 & 0.95 & $-45.0$ & $-0.02$ & 0.95& \\ 
&&$ S_{\frac{1}{2}}$ & $27.4$ & 0.18 & 0.88 & $-31.5$ & 0.35 & 0.85& \\ \hline \hline
\end{tabular*}
\caption{Parameterization of the transition form factors for $S_{11} (1535)$, $S_{11}(1650)$, $P_{11}(1710)$ resonances on the proton and neutron targets.  ${\cal A}_{\alpha} (0)$ is given in units of $10^{-3}$ GeV$^{-2}$ and the coefficients $a_1$  and $b_1$ in units of GeV$^{-2}$.}
\label{tab:resonance}
\end{table}

\begin{figure}
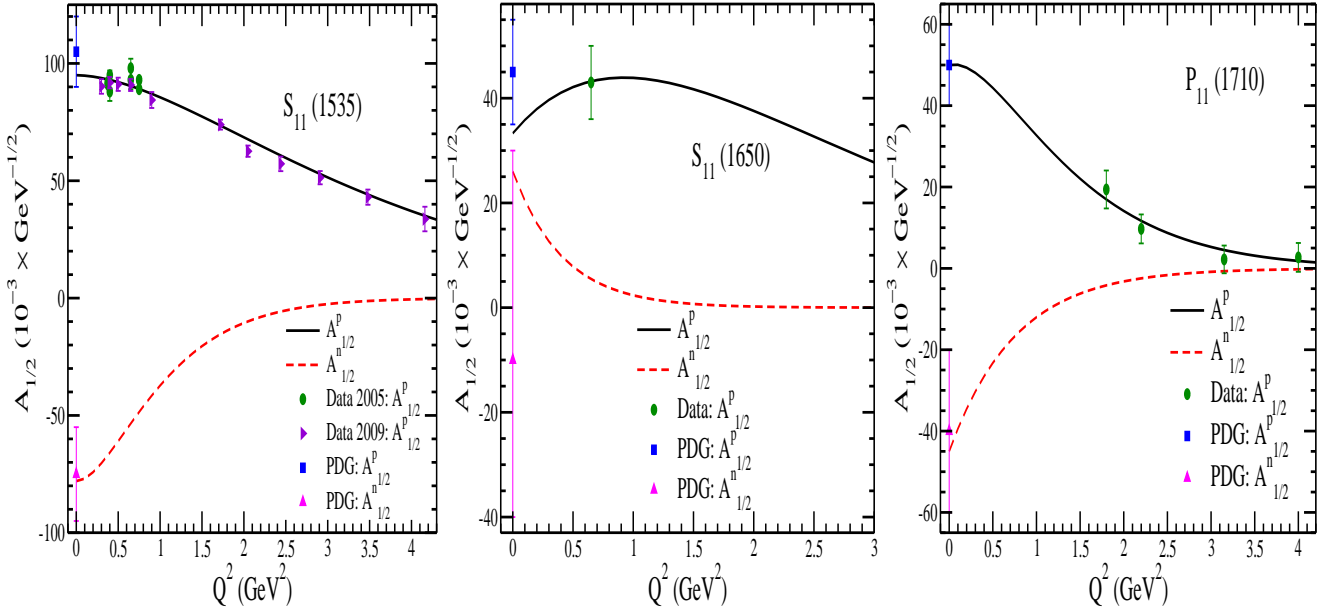

\begin{center}
\includegraphics[width=0.32\textwidth,height=.45\textwidth]{A12N_Q2_S11_1535.eps}
\includegraphics[width=0.32\textwidth,height=.45\textwidth]{A12N_Q2_S11_1650.eps}
\includegraphics[width=0.32\textwidth,height=.45\textwidth]{A12N_Q2_P11_1710.eps}
\caption{$Q^2$ dependence of the helicity amplitude $A_{1/2}^{p,n}$ appearing in Eq.~(\ref{eq:ffpar}) 
for $S_{11}(1535)$~(left panel), $S_{11}(1650)$~(middle panel), and $P_{11}(1710)$~(right panel) resonances. Solid~(dashed) lines are our results for $A_{1/2}^{p}~(A_{1/2}^{n})$ amplitudes. Solid square~(up triangle) are the values of $A_{1/2}^{p}~(A_{1/2}^{n})$ at $Q^2=0$ from the PDG~\cite{ParticleDataGroup:2020ssz}. Solid circle and right triangle are the data points available from the CLAS experiment~\cite{Aznauryan:2004jd, Aznauryan:2005tp, CLAS:2009ces, CLAS:2014fml}.}
\label{fg_eta:A12_S11}
\end{center}
\end{figure}
\begin{figure}
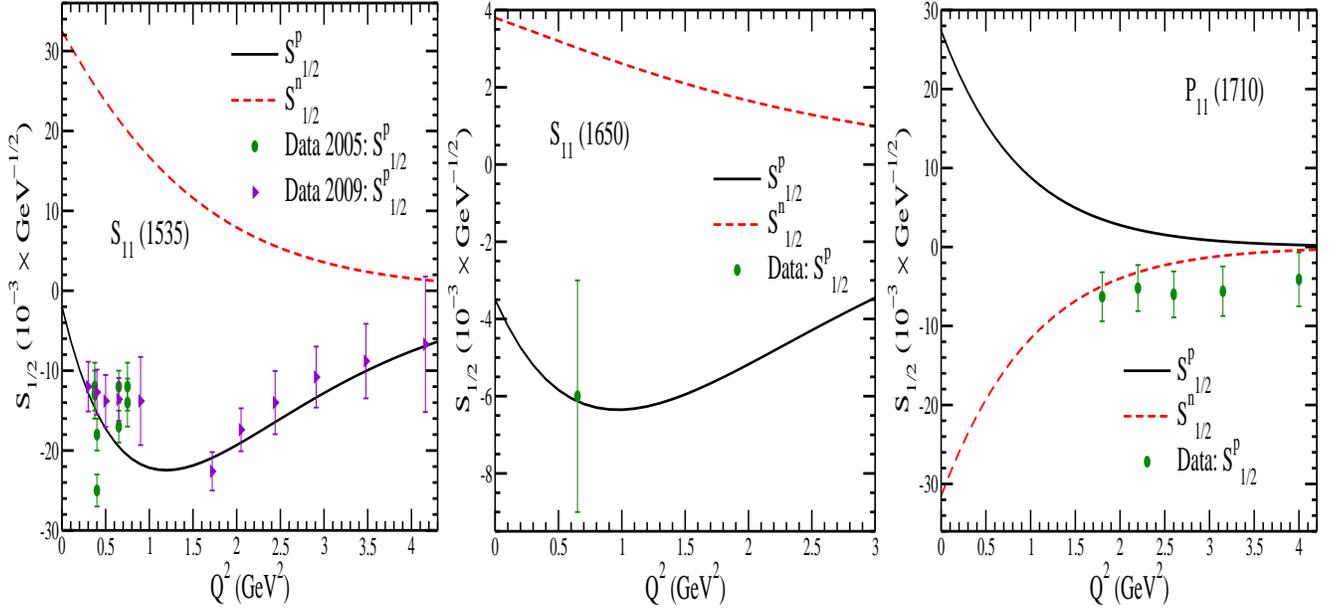

\begin{center}
\includegraphics[width=0.32\textwidth,height=.45\textwidth]{S12N_Q2_S11_1535.eps}
\includegraphics[width=0.32\textwidth,height=.45\textwidth]{S12N_Q2_S11_1650.eps}
\includegraphics[width=0.32\textwidth,height=.45\textwidth]{S12N_Q2_P11_1710.eps}
\caption{$Q^2$ dependence of the helicity amplitude $S_{1/2}^{p,n}$ given in Eq.~(\ref{eq:ffpar}) 
for $S_{11}(1535)$~(left panel), $S_{11}(1650)$~(middle panel), and $P_{11}(1710)$~(right panel) resonances. Lines and points have the same meaning as in Fig.~\ref{fg_eta:A12_S11}. }
\label{fg_eta:S12_S11}
\end{center}
\end{figure}
\begin{figure}
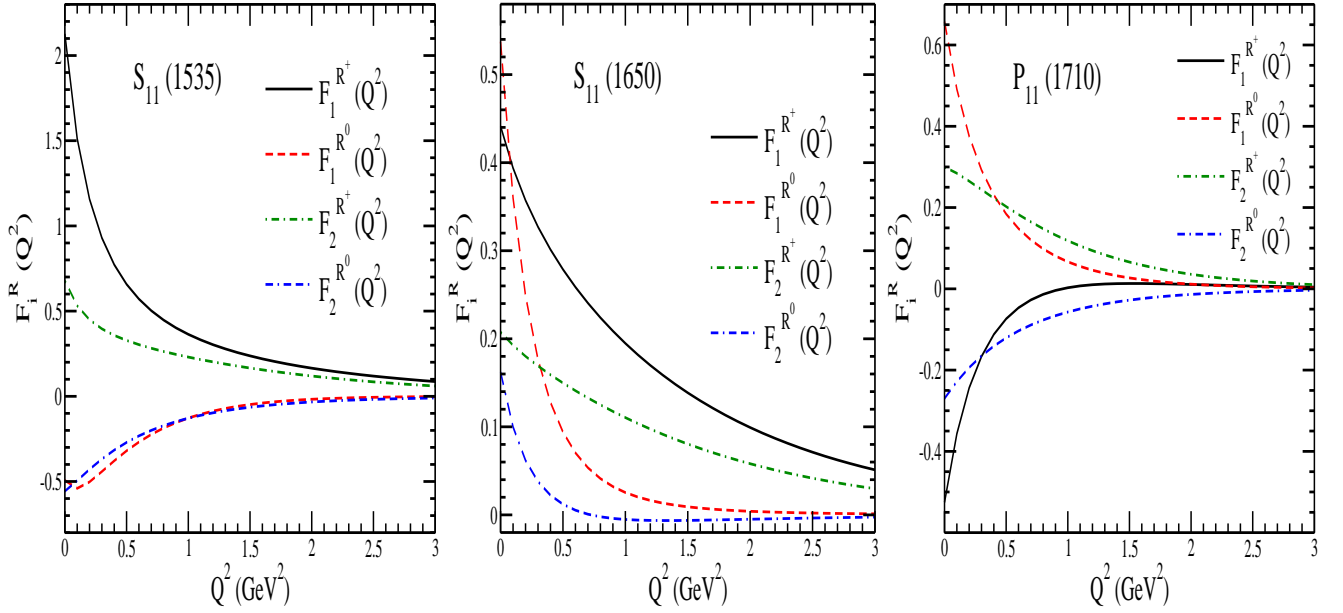

\begin{center}
\includegraphics[width=0.32\textwidth,height=.45\textwidth]{EM_FF_S11_1535.eps}
\includegraphics[width=0.32\textwidth,height=.45\textwidth]{EM_FF_S11_1650.eps}
\includegraphics[width=0.32\textwidth,height=.45\textwidth]{EM_FF_P11_1710.eps}
\caption{$Q^2$ dependence of the electromagnetic form factors $F_{i}^{R^+,R^0}(Q^2);~(i=1,2)$ given in Eq.~(\ref{eq:hel_em_ff}) 
for $S_{11}(1535)$~(left panel), $S_{11}(1650)$~(middle panel), and $P_{11}(1710)$~(right panel) resonances.  }
\label{fg_eta:ff_S11}
\end{center}
\end{figure}

Using Eq.~(\ref{eq1}) in Eq.~(\ref{eq2}), the helicity amplitudes $A_{\frac{1}{2}} (Q^2)$ and $S_{\frac{1}{2}} (Q^2)$ in terms of the electromagnetic
form factors
 $F_1^{R^+,R^0}$ and $F_2^{R^+,R^0}$ are obtained as~\cite{Leitner:2008ue}:
\begin{eqnarray}\label{eq:hel_em_ff}
A_{\frac{1}{2}}^{p,n} (Q^2)&=& \sqrt{\frac{2 \pi \alpha
}{M}\frac{(M_R+M)^2+Q^2}{M_R^2-M^2}} 
\left( \frac{Q^2}{4M^2} F_1^{R^+,R^0}(Q^2) + \frac{M_R-M}{2M} F_2^{R^+,R^0}(Q^2) \right) 
\nonumber \\
S_{\frac{1}{2}}^{p,n} (Q^2)&=& \sqrt{\frac{\pi \alpha }{M}\frac{(M_R - 
M)^2+Q^2}{M_R^2-M^2}} 
\frac{(M_R + M)^2+Q^2}{4 M_R M} \nonumber \\
&& \times \left(  \frac{M_R-M}{2M} F_1^{R^+,R^0}(Q^2) - F_2^{R^+,R^0}(Q^2) \right).
\end{eqnarray}
The $Q^2$ dependence of the helicity amplitudes~(Eq.~(\ref{eq:hel_em_ff}))  is generally parameterized as~\cite{Tiator:2011pw}:
\begin{equation}\label{eq:ffpar}
{\mathcal A}_{\alpha}(Q^2) = {\mathcal A}_{\alpha}(0) (1+\alpha Q^2)\, e^{-\beta Q^2} ,
\end{equation}
where $ {\mathcal A}_{\alpha}(Q^2)$ are the helicity amplitudes; $A_{\frac12}(Q^2)$ and $S_{\frac12}(Q^2)$ and parameters 
${\mathcal A}_{\alpha}(0)$ are generally determined by a fit to the photoproduction data of the corresponding resonance. 
In the present work, the values of $A_{\frac{1}{2}} (0)$ are taken from  the PDG~\cite{ParticleDataGroup:2020ssz}. 
While the parameters $\alpha$ and $\beta$ for each amplitude are obtained from the electroproduction data available at different 
$Q^2$ from the CLAS experiment~\cite{Aznauryan:2004jd, Aznauryan:2005tp, CLAS:2009ces, CLAS:2014fml}, and the values of these parameters for the different nucleon resonances parameterized are tabulated in Table~\ref{tab:resonance}.
The $Q^2$ dependence of the helicity amplitudes for $S_{11}(1535)$, $S_{11}(1650)$, and $P_{11}(1710)$ resonances is described by Eq.~(\ref{eq:ffpar}), which are then used in Eq.~(\ref{eq:hel_em_ff}) to obtain the electromagnetic $F_{1,2}^{R^+,R^0} (Q^2)$ form factors. 

In Figs.~\ref{fg_eta:A12_S11} and \ref{fg_eta:S12_S11}, we have shown the $Q^2$ dependence of  $A_{\frac{1}{2}}^{p,n}$ and $S_{\frac{1}{2}}^{p,n}$, respectively, for $S_{11} (1535)$, $S_{11} (1650)$, and $P_{11} (1710)$ resonances. The values of $A_{\frac{1}{2}}^{p,n} (0)$ are taken from PDG~\cite{ParticleDataGroup:2020ssz}. 
The data points shown in these figures are obtained 
from the analysis of the experimentally measured  differential cross sections, longitudinally polarized
beam asymmetries, and longitudinal target and beam-target asymmetries for $\pi$ and $\eta$ electroproductions from the proton target available from the CLAS experiment~\cite{Aznauryan:2004jd, Aznauryan:2005tp, CLAS:2009ces, CLAS:2014fml}.
To determine the helicity amplitudes at the different values of $Q^2$, in Refs.~\cite{Aznauryan:2004jd, Aznauryan:2005tp, CLAS:2009ces, CLAS:2014fml} the experimental data are analyzed using the two different approaches viz. dispersion relations and a unitary
isobar model, and the final results are the average of the two analyses.
The $Q^2$ dependence of the electromagnetic form factors $F_{i}^{R^+,R^0} (Q^2);~(i=1,2)$ for the resonances $S_{11}(1535)$, $S_{11}(1650)$, and $P_{11}(1710)$ are presented in Fig.~\ref{fg_eta:ff_S11}.

\begin{figure}
\begin{center}
\includegraphics[width=1.\textwidth,height=.8\textwidth]{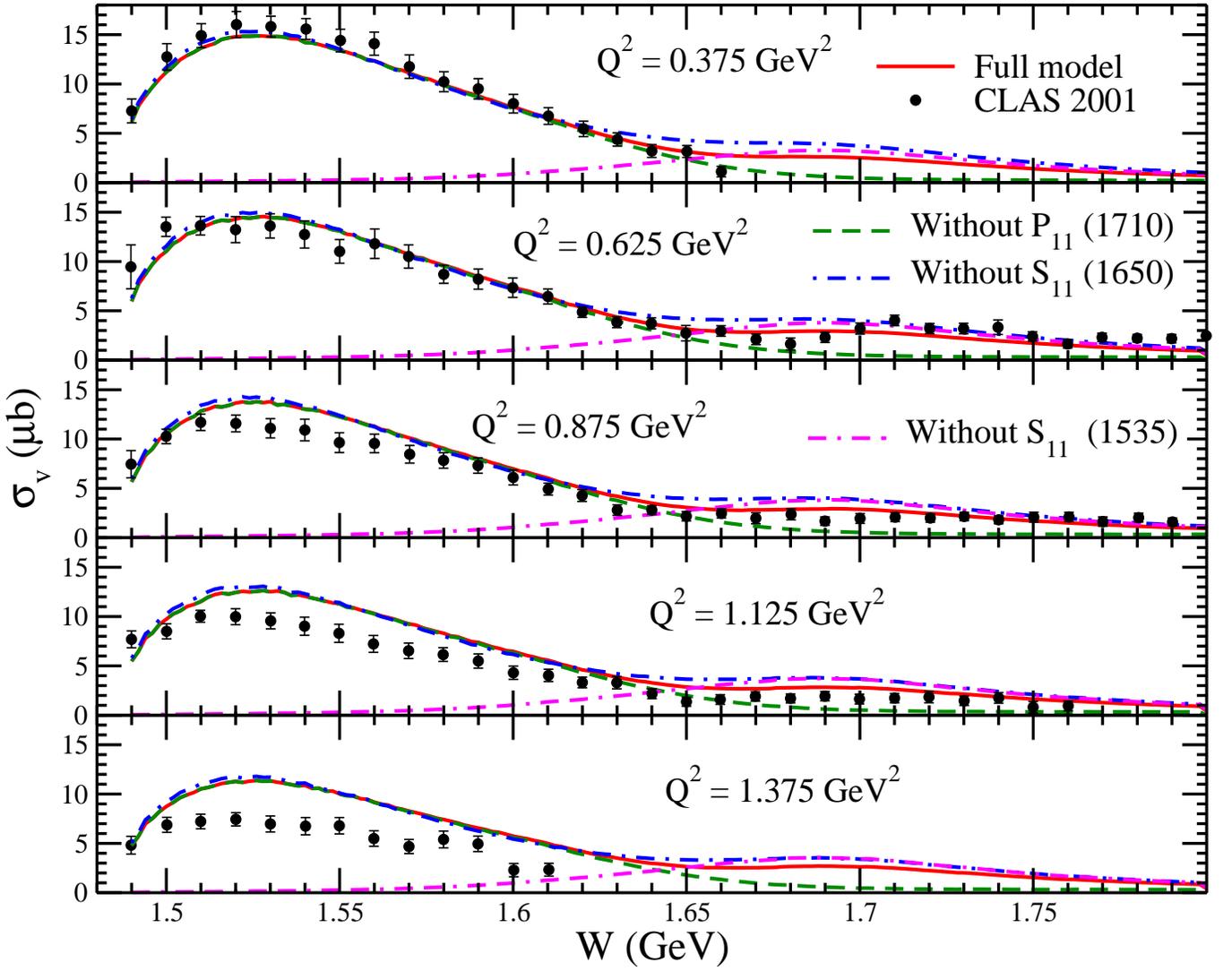}
\caption{Integrated cross section $\sigma_\text{v}$ vs $W$ at different $Q^2$ for $\gamma^\ast  p \rightarrow \eta p$ process. The experimental 
points are taken from the CLAS data~\cite{CLAS:2000mbw}. 
Solid line shows the results of the full model which receives contribution from the nonresonant Born terms as well as from the nucleon resonance excitations. Double-dashed-dotted, dashed-dotted, and dashed lines, respectively, show the results of the full model without $S_{11} (1535)$, $S_{11}(1650)$, and $P_{11}(1710)$ resonances.}
\label{fg_eta:electro_xsec_class}
\end{center}
\end{figure}
Using the expressions of the hadronic currents given in Eqs.~(\ref{Eq_eta:amp_photo}) and (\ref{jhad:eta_electro}), respectively, for the nonresonant Born terms and resonance excitations, and the leptonic current given in Eq.~(\ref{eq:lepton_current}) in Eq.~(\ref{eq:5fdxs}), we obtain the angular distribution~($\frac{d\sigma_\text{v}}{d\Omega_{qp_{\eta}}}$) of the scattering of virtual photons with proton. Performing integration over $\cos\theta_{qp_{\eta}}$ and $\phi_{qp_{\eta}}$, we obtain the total cross section $\sigma_\text{v}$ for $\gamma^\ast  p \rightarrow \eta p$ process, which is presented in Fig.~\ref{fg_eta:electro_xsec_class} as a function of CM energy $W$ at different values of $Q^2$ viz. $Q^2= 0.375$, 0.625, 0.875, 1.125, and 1.375 GeV$^2$. The theoretical calculations 
presented as  the full model receives contribution from the nonresonant Born terms as well as from the $S_{11}(1535)$, $S_{11} (1650)$, and 
$P_{11}(1710)$ resonances. In order to depict the individual contribution of the different nucleon resonances to the total cross section, we have also 
presented the results for the full model by switching off the contribution of an individual resonance. It may be observed from the figure that at all 
values of $Q^2$, the first peak region around $W\sim 1.52$~GeV is dominated by $S_{11}(1535)$ resonance while in the region of $W>1.65$~GeV, the relative
contribution from $P_{11}(1710)$ becomes significant. We have also compared our theoretical calculations with the experimental data available from the CLAS 
experiment~\cite{CLAS:2000mbw} and found a good agreement between the experimental and theoretical results especially in the low $Q^2$ region.

\subsection{$\eta$ production induced by (anti)neutrinos}\label{sec:eta:weak}
(Anti)neutrino induced single $\eta$ production off the nucleon target~(Fig.~\ref{Ch12_fg_eta:cc_weak_feynman}) are given by the 
following reactions
\begin{eqnarray}\label{Ch12_eq:eta_weak_process_cc}
\nu_\mu (k)  + n (p) &\longrightarrow& \mu^- (k^\prime) + \eta ( p_\eta) + p (p^\prime),   \\
\label{Ch12_eq:eta_weak_process_cc1}
\bar \nu_\mu (k)  + p (p) &\longrightarrow& \mu^+ (k^\prime) + \eta( p_\eta)  + n(p^\prime) ,
\end{eqnarray}
where the quantities in the parenthesis are the four momenta of the particles.
The general expression of the differential scattering cross section for the reactions shown in 
Eqs.~(\ref{Ch12_eq:eta_weak_process_cc}) and (\ref{Ch12_eq:eta_weak_process_cc1}) in the laboratory frame is given in 
Eq.~(\ref{eq:inelastic:reaction}), where transition matrix element, in the case of weak interaction induced process is given by
\begin{equation}\label{eq:mat:weak}
 {\cal M} = \frac{G_{F}}{\sqrt{2}} \cos\theta_{C}  l_{\mu} j^{\mu},
\end{equation}
where $G_{F}$ is the Fermi coupling constant and $\theta_{C}$ is the Cabibbo mixing angle.
The leptonic current $l_{\mu}$ is given 
\begin{equation}
 l_{\mu} = \bar{u} (k^{\prime}) \gamma_{\mu} (1-\gamma_5) u(k)
\end{equation}
and the hadronic current 
receives contribution from the nonresonant terms as well as from the resonance excitations and their subsequent 
decay to $N\eta$ final state.  
%
The hadronic currents for the Born diagrams~(s- and u-channels) with nucleon 
poles, using the nonlinear sigma model discussed in Section~\ref{sec:sigma_model}, are given in Eq.~(\ref{jhad:eta_electro}), except for the fact that ${\cal O}_{N}$ is now replaced by ${\cal O}_V$
where ${\cal O}_{V} = V^{\mu} -A^{\mu}$ is the weak vertex factor and $V^{\mu}$ and $A^{\mu}$ are defined 
in terms of the weak vector and axial-vector form factors as
\begin{align}  \label{eq:vectorspinhalfcurrent}
  V^{\mu}& ={f_{1}^{V}}(Q^2) \gamma^\mu + \frac{f_2^{V}(Q^2)}{2 M} 
  i \sigma^{\mu\nu} q_\nu ,  \\
    \label{eq:axialspinhalfcurrent}
  A^{\mu} &=  \left[{g_1}(Q^2) \gamma^\mu  +  \frac{g_3(Q^2)}{M} q^\mu\right] \gamma_5 ,
\end{align} 
where $f_{1,2}^V(Q^2)$ are respectively the isovector form factors, and $g_1(Q^2)$ and $g_3(Q^2)$ are 
the axial vector and pseudoscalar form factors. 
 The two isovector form factors 
$f_{1,2}^V(Q^2)$ are expressed in terms of the Dirac~($F_1^{p,n} (Q^2)$) and Pauli~($F_2^{p,n} (Q^2)$) form factors, discussed in Section~\ref{current_em}, for the 
protons and the neutrons  using the relationships,
\begin{equation}\label{Eq_eta:f1v_f2v}
f_{1,2}^V(Q^2)=F_{1,2}^p(Q^2)- F_{1,2}^n(Q^2). 
\end{equation}
These electromagnetic form factors may be rewritten in terms of Sachs' form factors using Eq.~(\ref{Eq_eta:f1pn_f2pn}).
 
The axial form factor, $g_1(Q^2)$ is parameterized as
\begin{equation}\label{Eq_eta:fa}
g_1(q^2)=g_A(0)~\left[1+\frac{Q^2}{M_A^2}\right]^{-2},
\end{equation}
where $g_A(0)=1.267$ is the axial charge and $M_A$ is the axial dipole mass, which in the numerical calculations is taken as the world average value i.e. $M_A = 1.026$~GeV~\cite{Bernard:2001rs}.
On the other hand pseudoscalar form factor $g_3(Q^2)$ may be expressed  in terms of $g_1(Q^2)$ 
using the PCAC hypothesis and Goldberger-Treiman relation as,
\begin{equation}\label{Eq:fp_nucleon}
g_3(Q^2)=\frac{2Mg_1(Q^2)}{m_\pi^2+Q^2},
\end{equation}
with $m_\pi$ as the pion mass.

Next, we discuss the positive and negative parity resonance excitation mechanism for the weak interaction induced $\eta$ production. 
The general expression of the hadronic current for the $s-$ and $u-$ channel resonance excitations and their subsequent decay to $N\eta$ mode are given in Eq.~\ref{jhad:eta_electro}, where the vertex factor $\Gamma_{\frac{1}{2}\pm}^{\mu}$ is now written as
\begin{align}\label{eq:vec_half_pos}
  \Gamma^{\mu}_{\frac{1}{2}^+} &= {V}^{\mu}_\frac{1}{2} - {A}^{\mu}_\frac{1}{2},
  \end{align}
  for the positive parity resonance, and as
\begin{align}\label{eq:vec_half_neg}
  \Gamma^{\mu}_{\frac{1}{2}^-} &= ({V}^{\mu}_\frac{1}{2} - {A}^{\mu}_\frac{1}{2}) \gamma_5 ,
  \end{align}
  for the negative parity resonance. The vector and axial vector vertex factors for the weak interaction processes are given by
\begin{align}  \label{eq:vectorspinhalfcurrent}
  V^{\mu}_{\frac{1}{2}} & =\frac{{f_{1}^{CC}}(Q^2)}{(2 M)^2}
  \left( Q^2 \gamma^\mu + {q\hspace{-.5em}/} q^\mu \right) + \frac{f_2^{CC}(Q^2)}{2 M} 
  i \sigma^{\mu\alpha} q_\alpha ,  \\
    \label{eq:axialspinhalfcurrent}
  A^{\mu}_\frac{1}{2} &=  \left[{g_1^{CC}}(Q^2) \gamma^\mu  +  \frac{g_3^{CC}(Q^2)}{M} q^\mu\right] \gamma_5 ,
\end{align}
where $f_i^{CC}(Q^2)$~($i=1,2$) are the isovector transition form factors which in turn are expressed in terms of the 
charged~($F_{i}^{R+} (Q^2)$) and neutral~($F_{i}^{R0} (Q^2)$) electromagnetic transition form factors as:
\begin{equation}\label{eq:f12vec_res_12}
f_i^{CC}(Q^2) = F_i^{R+}(Q^2) - F_i^{R0}(Q^2), \quad \quad i=1,2
\end{equation}
for isospin $\frac{1}{2}$ resonances. 
Further, these form factors are related to the helicity amplitudes as discussed in Section~\ref{current_em}.

The axial-vector current consists of two form factors viz. $g_1^{CC}(Q^2)$ and $g_3^{CC}(Q^2)$, which are determined 
assuming the PCAC hypothesis and pion pole dominance 
of the divergence of the axial-vector current through the generalized GT relation for $N 
- R$ transition~\cite{SajjadAthar:2022pjt}. The divergence of the axial vector current, defined in Eq.~(\ref{eq:axialspinhalfcurrent}), is obtained as
\begin{equation}\label{eq:g1_1}
 \partial_{\mu}A^{\mu}_{\frac{1}{2}} = \bar{u} (p_{R}) \left[g_1^{CC} (Q^2) (M_{R} \pm M) + \frac{g_3^{CC}}{M}q^2 \right] \gamma_5 \Gamma u(p)
\end{equation}
where $p_{R}$ is the four momentum of the resonance, $\Gamma = 1(\gamma_5)$ and $+(-)$ in $M_{R} \pm M$ stands for the positive~(negative) parity resonances. 

The Lagrangian at the strong $NR\pi$ vertex is written as
\begin{equation}\label{lag_pi}
 {\cal L}_{RN\pi} = \sqrt{2} \frac{g_{RN\pi}}{f_{\pi}} \bar{\Psi}_R \Gamma^\mu_s \partial_\mu \phi_i \tau^i \Psi
\end{equation}
where $g_{RN\pi}$ is the coupling constant at the strong vertex, $f_\pi$ is the pion decay constant, $\phi$ represents the triplet of the pion field. The value of the strong coupling $g_{RN\pi}$ is obtained using the partial decay width of the resonance to the $N\pi$ mode, using Eq.~(\ref{eq:12_width}) with ${\cal C}=3$ and $M_\eta$ is replaced by $m_{\pi}$.

The axial vector current obtained in the pion pole dominance of the divergence of the axial vector current~(using Eq.~(\ref{lag_pi})) is given by
\begin{equation}
A^{\mu}_{PP} = \sqrt{2} \bar{u}(p_{R}) \frac{g_{RN\pi}}{f_{\pi}} (M_R \pm M) \gamma_5\Gamma u (p) \frac{\sqrt{2} f_\pi q^\mu}{q^2-m_\pi^2}.
\end{equation}
The divergence of the above expression, in the limit of $m_\pi \rightarrow 0$, is obtained as
\begin{equation}
\partial_\mu A^{\mu}_{PP} = 2 g_{RN\pi} \bar{u}(p_{R}) (M_R \pm M) \gamma_5\Gamma u (p) .
\end{equation}
Comparing the term associated with $g_{1}^{CC} (Q^2)$ at $Q^2=0$ in Eq.~(\ref{eq:g1_1}) with the above expression, we obtain
\begin{equation}\label{eq:g1_pos}
g_1^{CC}(0)= 2 g_{RN\pi},
\end{equation}
with $g_{RN\pi}$ being the coupling strength for $R_{\frac12} \to N\pi$ decay, 
which has been determined by the partial decay width of the resonance and tabulated in Table~\ref{tab:param-p2}. Since no information about the $Q^2$ dependence of 
the axial-vector form factor is known experimentally, therefore, a dipole form is assumed as in the case of $N \rightarrow 
N^{\prime}$ or $N \rightarrow Y$ transitions:
\begin{equation}
 g_1^{CC}(Q^2) = \frac{g_1^{CC}(0)}{\left(1+\frac{Q^2}{M_{A}^2}\right)^2},
\end{equation}
with $M_{A}=1.026$~GeV, and the pseudoscalar form factor $g_3^{CC}(Q^2)$  is given by
\begin{equation}\label{eq:fp_res_spinhalf}
g_{3}^{CC}(Q^2) = \frac{(MM_{R}\pm M^{2})}{m_{\pi}^{2}+Q^{2}} g_1^{CC}(Q^2) ,
\end{equation}
where $+(-)$ sign is for positive~(negative) parity resonances. However, the contribution of $g_{3}^{CC} (Q^2)$ being directly proportional to the lepton mass squared is almost negligible.

\section{Results and Discussion}\label{results}
\begin{figure}[h]
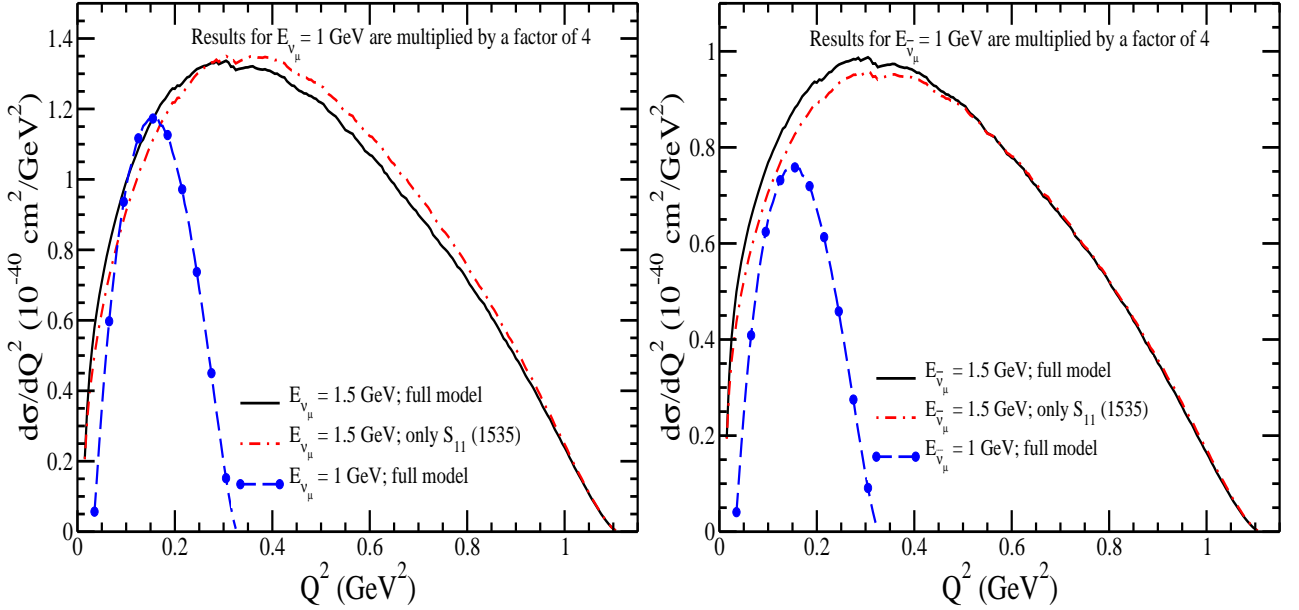

\begin{center}
\includegraphics[width=0.47\textwidth,height=.45\textwidth]{dsigma_dq2_neutrino.eps}
\includegraphics[width=0.47\textwidth,height=.45\textwidth]{dsigma_dq2_antineutrino.eps}
\caption{$Q^2$ distribution for the charged current induced $\nu_{\mu} + n \longrightarrow \mu^{-} + p + \eta$~(left panel) and $\bar{\nu}_{\mu} + p \longrightarrow \mu^{+} + n + \eta$~(right panel) processes
at $E_{\nu_\mu({\bar\nu}_\mu)}=1$~GeV~(dashed line with circle) and 1.5~GeV~(solid line) using the full model calculation. Dashed-dotted line shows the results obtained from $S_{11}(1535)$ resonance only at $E_{\nu_\mu({\bar\nu}_\mu)}=1.5$~GeV.}
\label{fg_eta:cc_q2_weak}
\end{center}
\end{figure}
\begin{figure}
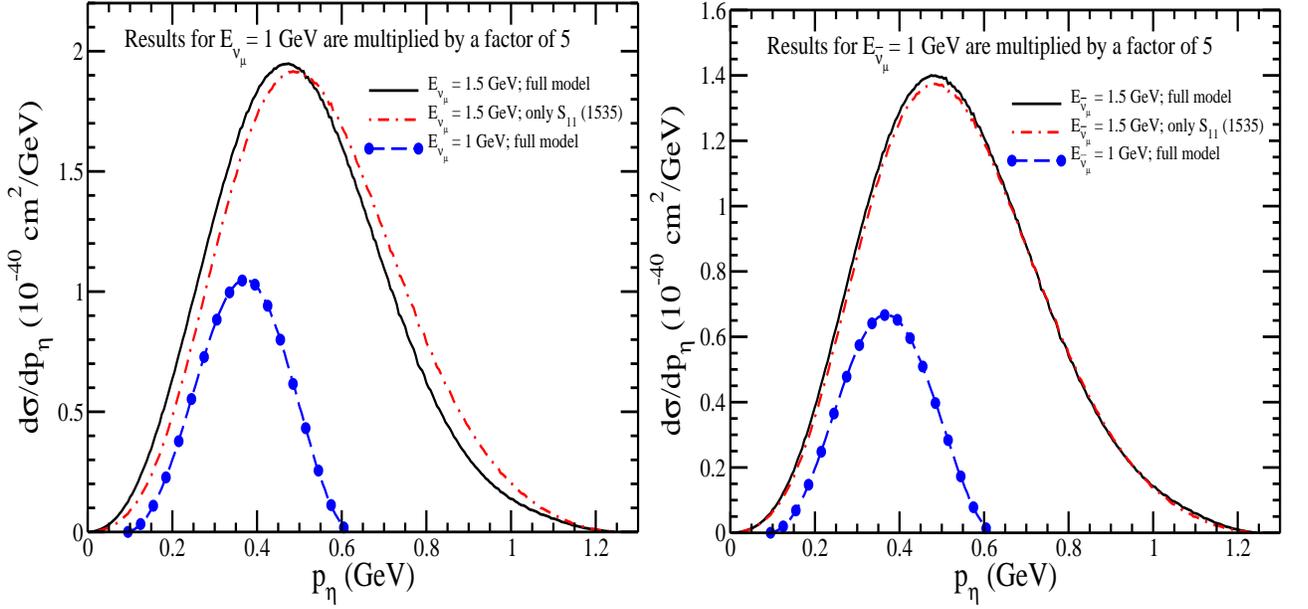

\begin{center}
\includegraphics[width=0.47\textwidth,height=.45\textwidth]{dsigma_dpeta_neutrino.eps}
\includegraphics[width=0.47\textwidth,height=.45\textwidth]{dsigma_dpeta_antineutrino.eps}
\caption{$\eta$-momentum distribution for the charged current induced $\nu_{\mu} + n \longrightarrow \mu^{-} + p + \eta$~(left panel) and $\bar{\nu}_{\mu} + p \longrightarrow \mu^{+} + n + \eta$~(right panel) processes. 
Lines and points have the same meaning as in Fig.~\ref{fg_eta:cc_q2_weak}.}
\label{fg_eta:cc_eta_weak}
\end{center}
\end{figure}
\begin{figure}
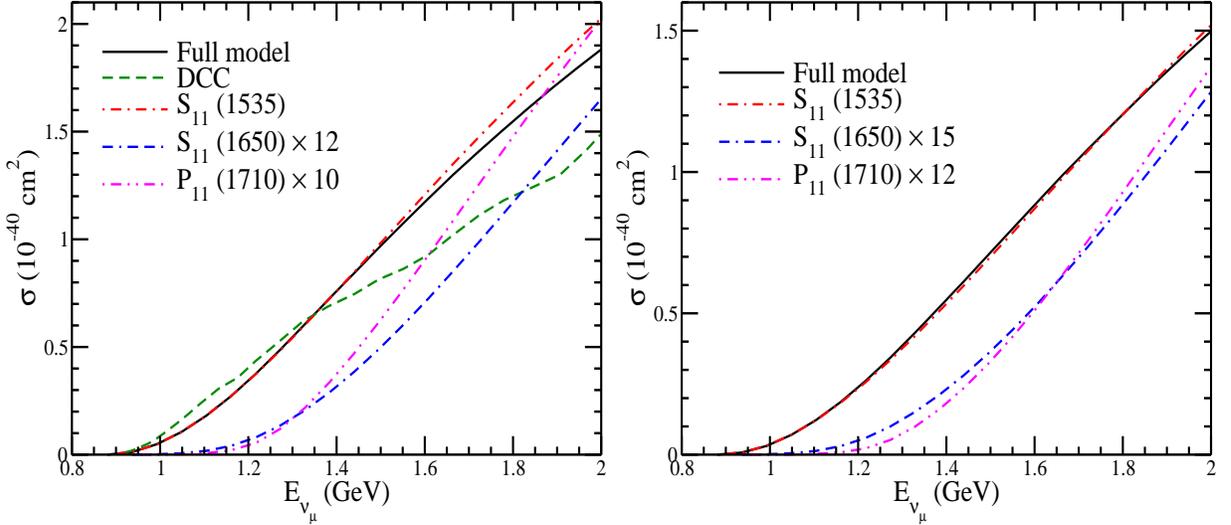

\begin{center}
\includegraphics[width=8cm,height=7cm]{neutrino_eta_resonance.eps}
\includegraphics[width=8cm,height=7cm]{antineutrino_eta_resonance.eps}
\caption{Total scattering cross section for CC induced $\eta$  production i.e. $\nu_{\mu} + n 
\longrightarrow \mu^{-} + \eta + p$~(left panel) and $\bar{\nu}_{\mu} + p \longrightarrow \mu^{+} + \eta + n$~(right panel). 
Full model~(solid line) consists of the contributions from all the diagrams including $S_{11}(1535)$, $S_{11}(1650)$, and $P_{11}(1710)$. The individual contribution from $S_{11}(1535)$, $S_{11}(1650)$, and $P_{11}(1710)$ resonances are shown by dashed-dotted, double-dashed-dotted, and double-dotted-dashed lines respectively.
In the case of neutrino induced $\eta$ production, we have also compared our results of the full model with the results 
obtained in the DCC model by Nakamura et al.~\cite{Nakamura:2015rta}.}\label{Ch12_fg_eta:cc_xsec_weak}
\end{center}
\end{figure}
\begin{figure}
\begin{center}
\includegraphics[width=8cm,height=7cm]{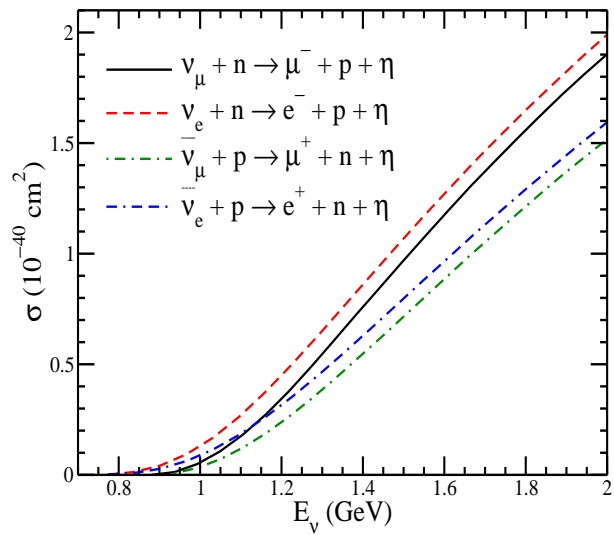}
\caption{Total scattering cross section for CC induced $\eta$  production i.e. $\nu_{\mu} + n 
\longrightarrow \mu^{-} + \eta + p$~(solid line), $\nu_{e} + n 
\longrightarrow e^{-} + \eta + p$~(dashed line), 
$\bar{\nu}_{\mu} + p \longrightarrow \mu^{+} + \eta + n$~(dashed-dotted line), and $\bar{\nu}_{e} + p \longrightarrow e^{+} + \eta + n$~(double-dashed-dotted line). }\label{Ch12_fg_eta:cc_xsec_emu}
\end{center}
\end{figure}

 In Fig.~\ref{fg_eta:cc_q2_weak}, we have presented the results for the $Q^2$ distribution in the charged current $\nu_{\mu}(\bar{\nu}_{\mu})$ induced $\eta$ production from the free nucleon target 
 at $E_{\nu_{\mu} (\bar{\nu}_{\mu})} =1$ and 1.5 GeV. It may be observed from the figure that at $E_{\nu}=1.5$~GeV the results obtained with only $S_{11}(1535)$ resonance are comparable to the results of the full model in the region of high $Q^2$ region. Even in the low $Q^2$ region, the contribution from the Born terms and other resonances is not more than 10\% of the total result.
 At $E_{\nu_{\mu} (\bar{\nu}_{\mu})} = 1$~GeV, we find that~(not shown here) the contribution of the higher resonances and Born terms is almost negligible as compared to the contribution of $S_{11}(1535)$ resonance. 

 In Fig.~\ref{fg_eta:cc_eta_weak}, we have presented the results for the $\eta$-momentum distribution of $\nu_{\mu}(\bar{\nu}_{\mu})$ induced $\eta$ production from the free nucleon target 
 at $E_{\nu_{\mu} (\bar{\nu}_{\mu})} =1$ and 1.5 GeV. As in the case of $Q^2$ distribution, we find very small contribution of the Born terms and higher resonances to the $p_{\eta}$ distribution as compared to the contribution of $S_{11}(1535)$ resonance.

Fig.~\ref{Ch12_fg_eta:cc_xsec_weak} shows the results for the total scattering cross sections for the processes $\nu_{\mu} + 
n \longrightarrow \mu^{-} + \eta + p$ and $\bar{\nu}_{\mu} + p \longrightarrow \mu^{+} + \eta + n$. The individual 
contributions from $S_{11}(1535)$ resonance excitations, where both the direct and crossed diagrams are considered, as well as 
the full model~(sum of all the diagrams) are shown. It may be observed from the figure that in the case of both neutrino and 
antineutrino induced reactions, $S_{11}(1535)$ has the dominant contribution followed by $P_{11}(1710)$ and $S_{11}(1650)$ resonances. We have also compared the present results for the neutrino induced $\eta$ production 
with the results of DCC model~\cite{Nakamura:2015rta} and found that from threshold up to $E_{\nu_{\mu}} \sim 1.3$~GeV our 
results are consistent with the results of DCC model. While for $E_{\nu_{\mu}} > 1.3$~GeV, our results are higher than the 
results obtained using DCC model. 

To explicitly show the lepton mass effect, in Fig.~\ref{Ch12_fg_eta:cc_xsec_emu}, we have presented the results for the total scattering cross section for the $\eta$ production induced by $\nu_{e}(\bar{\nu}_{e})$ and $\nu_{\mu}(\bar{\nu}_{\mu})$. It may be noticed that due to the lower threshold for $\nu_{e}(\bar{\nu}_{e})$, the cross section is larger than that of $\nu_{\mu} (\bar{\nu}_{\mu})$ induced processes even at $E_{\nu(\bar{\nu})} = 2$~GeV. Since the value of axial dipole mass is in debate as the different neutrino experiment quote different values of $M_{A}$ ranging from 1 to 1.3~GeV, therefore, in the case of $\eta$ production, we have studied the effect of $M_{A}$ on the total scattering cross section~(not shown here) by varying it in the range $\pm 10\%$ for the Born as well as the resonance terms. We find that the effect of $M_{A}$ variation is less than 5\%, even at $E_{\nu_{\mu}(\bar{\nu}_{\mu})}=2$~GeV.

\section{Summary and conclusions}\label{conclude}
 In summary, we have studied the charged current $\nu_\mu({\bar\nu}_\mu)$  induced $\eta$ production off the nucleons and presented the results for the total scattering cross section $\sigma(E_{\nu_{\mu}(\bar{\nu}_{\mu})})$, $Q^2$-distribution~$\left(\frac{d\sigma}{dQ^2}\right)$ and the momentum distribution~$\left(\frac{d\sigma}{dp_\eta}\right)$ for the $\eta$ mesons, in a model in which the contribution from the nonresonant Born terms and the resonant terms are calculated in an effective Lagrangian approach. 
 The parameters of the model for evaluating the vector current contribution have been fixed by fitting the experimental data on the total cross sections of the photo- and electro- production on the nucleons from the MAINZ and CLAS experiments. 
 The vector part of the N $\longrightarrow R$ transition form factors has been obtained from the MAID helicity amplitudes while the axial vector part is obtained with the help of PCAC hypothesis and assuming the pion-pole dominance of the divergence of the axial vector current. 
 For the nonresonant background terms we have used a microscopical model based on the SU(3) chiral Lagrangians. The parameters of the model are  meson decay constants, Cabibbo’s angle, the proton and neutron magnetic moments, all of which are known with good accuracy, and the parameters $D$ and $F$ obtained from the analysis of the hyperon semileptonic decays. 

The results are summarized as
\begin{itemize}
 \item [(i)] In the electromagnetic sector, the cross section is dominated by the contribution from $S_{11}(1535)$ resonance, especially at low $W$. With the increase in $W$ the contribution from $P_{11}(1710)$ starts to show up and around $W \ge 1.7$~GeV its contribution is larger than the other resonances.
 
 \item [(ii)] In the weak sector, $S_{11}(1535)$ resonance dominates even at $E_{\nu} = 2$~GeV, while the contribution of $S_{11}(1650)$ is not more than 7\% of the total scattering cross section, while that of $P_{11}(1710)$ is about 10\%.
 
 \item [(iii)] The $Q^2$ distribution has a peak at $Q^2 \approx 0.15~(0.28)$~GeV$^2$ at $E_{\nu_{\mu}}=1~(1.5)$GeV. The nature of $Q^2$ distribution in the antineutrino channel is almost similar to that observed in the neutrino case.
 
 \item [(iv)] The eta momentum distribution has a peak at $p_{\eta} \approx 0.37~(0.47)$~GeV at $E_{\nu_{\mu}}=1~(1.5)$GeV. The nature of momentum distribution in the antineutrino channel is almost similar to that observed in the neutrino case.
\end{itemize}

To conclude, the results obtained in this work may be very useful in studying the feasibility of observing the weak production of $\eta$ in experiments like MicroBooNE, T2K, NOvA, MINERvA and the next generation experiments like T2-HyperK and DUNE.  Furthermore, the study of $\eta$ production is also important in the analysis of atmospheric neutrino events.

\section*{Acknowledgements}
MSA is thankful to M. J. Vicente Vacas, L. Alvarez Ruso, and M. Rafi Alam for many useful discussions. MSA and AF are thankful to the
Department of Science and Technology (DST), Government of India for providing financial assistance under Grant No.
SR/MF/PS-01/2016-AMU.

 \bibliographystyle{apsrev}

\end{document}